\documentclass[12pt]{article}

\usepackage{axodraw}

\makeatletter

\def\paragraph{\@startsection{paragraph}{4}{\z@}{+2.00ex plus
 +1ex minus +.2ex}{1.5ex plus .2ex}{\it\normalsize}}

\def\section{\@startsection {section}{1}{\z@}{+3.0ex plus +1ex minus
  +.2ex}{2.3ex plus .2ex}{\normalsize\bf\boldmath}}
\def\subsection{\@startsection{subsection}{2}{\z@}{+2.5ex plus +1ex
minus +.2ex}{1.5ex plus .2ex}{\normalsize\bf\boldmath}}
\def\subsubsection{\@startsection{subsubsection}{3}{\z@}{+3.25ex plus
 +1ex minus +.2ex}{1.5ex plus .2ex}{\normalsize\it}}

 \expandafter\ifx\csname mathrm\endcsname\relax\def\mathrm#1{{\rm #1}}\fi

\renewcommand{\theequation}{\thesection.\arabic{equation}}
\newcounter{saveeqn}

\@addtoreset{equation}{section}

\newcount\@tempcntc
\def\@citex[#1]#2{\if@filesw\immediate\write\@auxout{\string\citation{#2}}\fi
  \@tempcnta\z@\@tempcntb\m@ne\def\@citea{}\@cite{\@for\@citeb:=#2\do
    {\@ifundefined
       {b@\@citeb}{\@citeo\@tempcntb\m@ne\@citea
        \def\@citea{,\penalty\@m\ }{\bf ?}\@warning
       {Citation `\@citeb' on page \thepage \space undefined}}%
    {\setbox\z@\hbox{\global\@tempcntc0\csname
b@\@citeb\endcsname\relax}%
     \ifnum\@tempcntc=\z@ \@citeo\@tempcntb\m@ne
       \@citea\def\@citea{,\penalty\@m}
       \hbox{\csname b@\@citeb\endcsname}%
     \else
      \advance\@tempcntb\@ne
      \ifnum\@tempcntb=\@tempcntc
      \else\advance\@tempcntb\m@ne\@citeo
      \@tempcnta\@tempcntc\@tempcntb\@tempcntc\fi\fi}}\@citeo}{#1}}

\def\@citeo{\ifnum\@tempcnta>\@tempcntb\else\@citea
  \def\@citea{,\penalty\@m}%
  \ifnum\@tempcnta=\@tempcntb\the\@tempcnta\else
   {\advance\@tempcnta\@ne\ifnum\@tempcnta=\@tempcntb \else
\def\@citea{--}\fi
    \advance\@tempcnta\m@ne\the\@tempcnta\@citea\the\@tempcntb}\fi\fi}

\newcommand{\lsim}
{\mathrel{\raisebox{-.3em}{$\stackrel{\displaystyle <}{\sim}$}}}
\newcommand{\gsim}
{\mathrel{\raisebox{-.3em}{$\stackrel{\displaystyle >}{\sim}$}}}
\def\asymp#1%
{\mathrel{\raisebox{-.4em}{$\widetilde{\scriptstyle #1}$}}}

\def\Nequal#1%
{\mathrel{\raisebox{-.5em}{$\stackrel{=}{\scriptstyle\rm#1}$}}}
\newcommand{\dsl}[1]{\not \hspace{-0.7mm}#1}
\def\dsl{\mathpalette\make@slash}
\def\make@slash#1#2{\setbox\z@\hbox{$#1#2$}%
  \hbox to 0pt{\hss$#1/$\hss\kern-\wd0}\box0}

\def\beq{\begin{equation}}
\def\eeq{\end{equation}}
\def\beqar{\begin{eqnarray}}
\def\eeqar{\end{eqnarray}}
\def\barr#1{\begin{array}{#1}}
\def\earr{\end{array}}
\def\bfi{\begin{figure}}
\def\efi{\end{figure}}
\def\btab{\begin{table}}
\def\etab{\end{table}}
\def\bce{\begin{center}}
\def\ece{\end{center}}
\def\nn{\nonumber}

\def\text{\textstyle}

\def\al{\alpha}

\def\ga{\gamma}
\def\de{\delta}
\def\De{\Delta}
\def\eps{\epsilon}
\def\veps{\varepsilon}

\def\la{\lambda}
\def\om{\omega}

\def\si{\sigma}

\def\refeq#1{\mbox{(\ref{#1})}}

\def\reffi#1{\mbox{Figure~\ref{#1}}}
\def\reffis#1{\mbox{Figures~\ref{#1}}}
\def\refta#1{\mbox{Table~\ref{#1}}}

\def\refse#1{\mbox{Section~\ref{#1}}}

\def\refapp#1{\mbox{App.~\ref{#1}}}
\def\citere#1{\mbox{Ref.~\cite{#1}}}
\def\citeres#1{\mbox{Refs.~\cite{#1}}}

\newcommand{\TeV}{\unskip\,\mathrm{TeV}}
\newcommand{\GeV}{\unskip\,\mathrm{GeV}}

\newcommand{\pba}{\unskip\,\mathrm{pb}}
\newcommand{\fb}{\unskip\,\mathrm{fb}}

\newcommand{\ri}{{\mathrm{i}}}
\newcommand{\rd}{{\mathrm{d}}}

\newcommand{\Ord}{\mathswitch{{\cal{O}}}}
\newcommand{\Oa}{\mathswitch{{\cal{O}}(\alpha)}}

\renewcommand{\L}{{\cal{L}}}
\newcommand{\M}{{\cal{M}}}
\def\mathswitchr#1{\relax\ifmmode{\mathrm{#1}}\else$\mathrm{#1}$\fi}

\newcommand{\PW}{\mathswitchr W}
\newcommand{\Pw}{\mathswitchr w}
\newcommand{\PZ}{\mathswitchr Z}

\newcommand{\PH}{\mathswitchr H}
\newcommand{\Pe}{\mathswitchr e}
\newcommand{\Pne}{\mathswitch \nu_{\mathrm{e}}}
\newcommand{\Pnebar}{\mathswitch \bar\nu_{\mathrm{e}}}
\newcommand{\Pnmu}{\mathswitch \nu_{\mu}}
\newcommand{\Pnmubar}{\mathswitch \bar\nu_{\mu}}
\newcommand{\Pd}{\mathswitchr d}
\newcommand{\Pdbar}{\bar{\mathswitchr d}}
\newcommand{\Pu}{\mathswitchr u}
\newcommand{\Pubar}{\bar{\mathswitchr u}}
\newcommand{\Ps}{\mathswitchr s}

\newcommand{\Pc}{\mathswitchr c}
\newcommand{\Pcbar}{\bar{\mathswitchr c}}
\newcommand{\Pb}{\mathswitchr b}

\newcommand{\Pt}{\mathswitchr t}

\newcommand{\Pep}{\mathswitchr {e^+}}
\newcommand{\Pem}{\mathswitchr {e^-}}

\newcommand{\Pmup}{\mathswitchr {\mu^+}}
\newcommand{\Pmum}{\mathswitchr {\mu^-}}

\def\mathswitch#1{\relax\ifmmode#1\else$#1$\fi}

\newcommand{\MW}{\mathswitch {M_\PW}}

\newcommand{\MZ}{\mathswitch {M_\PZ}}
\newcommand{\MH}{\mathswitch {M_\PH}}
\newcommand{\Me}{\mathswitch {m_\Pe}}
\newcommand{\Mmy}{\mathswitch {m_\mu}}
\newcommand{\Mta}{\mathswitch {m_\tau}}
\newcommand{\Md}{\mathswitch {m_\Pd}}
\newcommand{\Mu}{\mathswitch {m_\Pu}}
\newcommand{\Ms}{\mathswitch {m_\Ps}}
\newcommand{\Mc}{\mathswitch {m_\Pc}}
\newcommand{\Mb}{\mathswitch {m_\Pb}}
\newcommand{\Mt}{\mathswitch {m_\Pt}}
\newcommand{\GW}{\Gamma_{\PW}}
\newcommand{\GZ}{\Gamma_{\PZ}}

\newcommand{\GH}{\Gamma_{\PH}}

\newcommand{\sw}{\mathswitch {s_\Pw}}
\newcommand{\cw}{\mathswitch {c_\Pw}}

\newcommand{\GF}{\mathswitch {G_\mu}}

\def\solid{\raise.9mm\hbox{\protect\rule{1.1cm}{.2mm}}}
\def\dash{\raise.9mm\hbox{\protect\rule{2mm}{.2mm}}\hspace*{1mm}}

\newcommand{\Born}{{\mathrm{Born}}}
\newcommand{\DPA}{{\mathrm{DPA}}}
\newcommand{\IBA}{{\mathrm{IBA}}}

\newcommand{\virt}{{\mathrm{virt}}}
\newcommand{\soft}{{\mathrm{soft}}}
\newcommand{\coll}{{\mathrm{coll}}}
\newcommand{\sub}{{\mathrm{sub}}}
\newcommand{\sing}{{\mathrm{sing}}}
\newcommand{\finite}{{\mathrm{finite}}}

\newcommand{\ffp}{\mathswitch{\mathrm{f\/f}'}}
\newcommand{\mfp}{\mathswitch{\mathrm{mf}'}}
\newcommand{\mmp}{\mathswitch{\mathrm{mm}'}}
\def\sub{{\mathrm{sub}}}
\def\gsub{g^{(\sub)}}
\def\Gsub{G^{(\sub)}}

\def\bcGsub{\bar {\cal G}^{(\sub)}}

\def\Li{\mathop{\mathrm{Li}_2}\nolimits}

\def\Re{\mathop{\mathrm{Re}}\nolimits}
\def\Im{\mathop{\mathrm{Im}}\nolimits}

\def\lra{\mathop{\mathrm{\leftrightarrow}}\nolimits}

\newcommand{\ggww}{\gamma\gamma\PW^+\PW^-}

\newcommand{\ggffff}{\gamma\gamma\to 4f}
\newcommand{\eeffff}{\Pep\Pem\to 4f}
\newcommand{\eewwffff}{\Pep\Pem\to\PW\PW\to 4f}

\newcommand{\ggffffg}{\ggffff\ga}

\newcommand{\ggffffkg}{\ggffff(\ga)}
\newcommand{\ggtoww}{\gamma\gamma\to \PW\PW}

\newcommand{\ggwwffff}{\gamma\gamma\to \PW\PW\to 4f}

\newcommand{\Whizard}{{\sc Whizard}}
\newcommand{\Madgraph}{{\sc Madgraph}}
\newcommand{\HDECAY}{{\sc HDECAY}}
\newcommand{\CompAZ}{{\sc CompAZ}}

\def\draftdate{\relax}
\def\mda{\relax}
\def\mua{\relax}
\def\mla{\relax}
\def\draft{
\def\thtystars{******************************}
\def\sixtystars{\thtystars\thtystars}
\typeout{}
\typeout{\sixtystars**}
\typeout{* Draft mode!
         For final version remove \protect\draft\space in source file *}
\typeout{\sixtystars**}
\typeout{}
\def\draftdate{\today}
\def\mua{\marginpar[\boldmath\hfil$\uparrow$]%
                   {\boldmath$\uparrow$\hfil}%
                    \typeout{marginpar: $\uparrow$}\ignorespaces}
\def\mda{\marginpar[\boldmath\hfil$\downarrow$]%
                   {\boldmath$\downarrow$\hfil}%
                    \typeout{marginpar: $\downarrow$}\ignorespaces}
\def\mla{\marginpar[\boldmath\hfil$\rightarrow$]%
                   {\boldmath$\leftarrow $\hfil}%
                    \typeout{marginpar: $\lra$}\ignorespaces}
\def\Mua{\marginpar[\boldmath\hfil$\Uparrow$]%
                   {\boldmath$\Uparrow$\hfil}%
                    \typeout{marginpar: $\uparrow$}\ignorespaces}
\def\Mda{\marginpar[\boldmath\hfil$\Downarrow$]%
                   {\boldmath$\Downarrow$\hfil}%
                    \typeout{marginpar: $\downarrow$}\ignorespaces}
\def\Mla{\marginpar[\boldmath\hfil$\Rightarrow$]%
                   {\boldmath$\Leftarrow $\hfil}%
                    \typeout{marginpar: $\lra$}\ignorespaces}
\overfullrule 5pt
\oddsidemargin -15mm
\marginparwidth 29mm
}

\def\stars{\strut\leaders\hbox{*}\hfill\strut}
\def\starline{\hfil\strut\hfil\hbox to \textwidth {\stars}\hfil}

\begin{document}
\thispagestyle{empty}
\def\thefootnote{\fnsymbol{footnote}}
\setcounter{footnote}{1}
\null
\draftdate\hfill MPP-2005-24\\
\strut\hfill hep-ph/0506005\\
\vspace{1.5cm}
\begin{center}
{\Large \bf\boldmath
Four-fermion production at $\ga\ga$ colliders:
\\[.5em]
2.\ Radiative corrections in double-pole approximation
\par} \vskip 2.5em
\vspace{1cm}

{\large
{\sc A.\ Bredenstein, S.\ Dittmaier and M. Roth} } \\[1cm]

{\it Max-Planck-Institut f\"ur Physik (Werner-Heisenberg-Institut)\\
D-80805 M\"unchen, Germany} \\
\par \vskip 1em
\end{center}\par
\vfill
{\bf Abstract:} \par
The ${\cal O}(\al)$ electroweak radiative corrections to $\ggwwffff$ within 
the electroweak Standard Model are calculated
in double-pole approximation (DPA). 
Virtual corrections are treated in DPA, leading to a
classification into factorizable and non-factorizable contributions,
and real-photonic corrections are based on complete lowest-order
matrix elements for $\ggffff{+}\ga$. Soft and collinear 
singularities appearing in the virtual and real corrections are combined
alternatively in two different ways, namely by using the dipole 
subtraction method or by applying phase-space slicing.  
The radiative corrections are implemented in a Monte Carlo generator
called {\sc Coffer$\gamma\gamma$}%
\footnote{The computer code can be obtained from the authors upon request.}%
, which optionally includes anomalous
triple and quartic gauge-boson couplings in addition and performs a
convolution over realistic spectra of the photon beams.
A detailed survey of numerical results comprises 
${\cal O}(\al)$ corrections to
integrated cross sections as well as to
angular, energy, and invariant-mass distributions.
Particular attention is paid to the issue of collinear-safety in the
observables.
\par
\vskip 1cm
\noindent
June 2005
\null
\setcounter{page}{0}
\clearpage
\def\thefootnote{\arabic{footnote}}
\setcounter{footnote}{0}

\section{Introduction}
\label{se:intro}

As an option at a future $\Pep\Pem$ linear collider,
a photon (or $\ga\ga$) collider \cite{Ginzburg:1981vm} 
found considerable interest in recent years.
It could provide us with information about new physics phenomena,
such as properties of Higgs bosons or of new particles,
which is in many respects complementary in the
$\Pep\Pem$ and $\ga\ga$ modes
(see, e.g., \citeres{Ginzburg:1981vm,DeRoeck:2003gv} and references therein).
Moreover, a $\ga\ga$ collider is a true W-boson-pair factory,
owing to the extremely high W-pair cross section, which tends to a
constant of about $80\pba$ in the high-energy limit (in the absence
of phase-space cuts), opening the possibility of precision studies in the
sector of electroweak gauge bosons. Either way, whether one is
interested in W-boson precision physics or in the search for new phenomena,
precise predictions for W-pair production are indispensable for signal
and background studies.

In our previous work \cite{Bredenstein:2004ef} we have made the first step
towards a precision calculation for the processes $\ggwwffff(+\ga)$ by
constructing a Monte Carlo event generator for lowest-order predictions
based on complete matrix elements for the processes $\ggffff$ and
$\ggffffg$. The possibility to convolute the cross sections with
realistic photon beam spectra is offered upon using the
parametrization of {\CompAZ} \cite{Zarnecki:2002qr}.
The Standard Model (SM) predictions were successfully compared
to results obtained with the multi-purpose packages 
{\Whizard} \cite{Kilian:2001qz} and {\Madgraph} \cite{Stelzer:1994ta}.
Moreover, we included an effective $\ga\ga\PH$ coupling, which is induced
by loop diagrams, 
as well as anomalous triple and quartic gauge-boson couplings.
The former is needed for studying Higgs 
production in the $s$-channel. An analysis of anomalous gauge-boson couplings
in $\ggtoww$ provides direct information on the $\gamma\PW\PW$ and 
$\gamma\gamma\PW\PW$ interactions without interference from the 
$\PZ$-boson sector. Both the Higgs resonance in the $s$-channel and the
more direct access to the $\gamma\PW\PW$ and
$\gamma\gamma\PW\PW$ interactions are complementary to the
situation in $\Pep\Pem$ annihilation.

In this paper we extend our lowest-order calculation \cite{Bredenstein:2004ef}
for $\ggffff$ by including the electroweak radiative corrections of
${\cal O}(\alpha)$ to the W-pair channels $\ggwwffff$
in the so-called ``double-pole approximation'' (DPA). 
The DPA extracts those contributions of the ${\cal O}(\alpha)$
corrections that are enhanced by two resonant W-boson
propagators, i.e.\ it represents the leading term in an expansion of
the cross section about the two W-propagator poles.
Note that tree-level diagrams for $\ggffff$ with at most one
resonant W~boson are suppressed w.r.t.\ the doubly-resonant 
$\ggtoww$ signal by a factor of ${\cal O}(\GW/\MW)\sim {\cal O}(\alpha)$.
Consequently, predictions based on full lowest-order matrix elements for
$\ggffff$ and ${\cal O}(\alpha)$ corrections for $\ggwwffff$ in DPA
should be precise up to terms of ${\cal O}(\alpha/\pi\times \GW/\MW)$,
since corrections typically involve the factor $\alpha/\pi$.
Including a quite conservative numerical safety factor, the relative
uncertainty should thus be $\lsim0.5\%$ for such predictions,
as long as neglected effects are not additionally enhanced.
The naive error estimate can, in particular, be spoiled by the occurrence of
large scale ratios, which exist, e.g., near production thresholds or
at very high energies. The estimate has recently been confirmed
for $\eewwffff$ with centre-of-mass (CM) energies
$170\GeV\lsim\sqrt{s}\lsim300\GeV$ by comparing a full
${\cal O}(\alpha)$ calculation \cite{Denner:2005es,Denner:2005fg}
with the corresponding DPA predictions provided by {\sc RacoonWW}
\cite{Denner:1999gp,Denner:2000bj}.

In detail, we apply the DPA only to the virtual corrections to
$\ggwwffff$, while we base the real-photonic corrections on complete
lowest-order matrix elements for $\ggffffg$. Apart from the treatment
of IR (soft and collinear) singularities, we can use 
the calculation of the bremsstrahlung
processes $\ggffffg$ for massless fermions
described in \citere{Bredenstein:2004ef}.
The concept of the DPA was already described in \citere{Aeppli:1993rs} 
for the corrections to $\eewwffff$ and later successfully
applied to these processes in different versions
\cite{Denner:1999gp,Denner:2000bj,Jadach:1998tz,Beenakker:1998gr,%
Kurihara:2001um}.
We follow the strategy of {\sc RacoonWW} \cite{Denner:1999gp,Denner:2000bj}
and adapt it to $\gamma\gamma$ collisions where necessary.
The virtual corrections in DPA can be naturally split into
factorizable and non-factorizable contributions.
The former comprise the corrections to on-shell W-pair
production \cite{Denner:1994ca,Denner:1995jv,Jikia:1996uu}%
\footnote{Radiative corrections to on-shell W-pair production,
$\ga\ga\to\PW\PW$, were also considered in \citere{Marfin:2003yq}.}
and the decay \cite{Bardin:1986fi} of on-shell W~bosons.
The latter account for soft-photon exchange between the production
and decay subprocesses; the known results for the non-factorizable
corrections \cite{Melnikov:1995fx,Denner:1997ia}
for $\eewwffff$ can be taken over to $\gamma\gamma$ collisions with
minor modifications.
Although the basic building blocks for the virtual corrections exist
in the literature, the combination into a complete set of
${\cal O}(\alpha)$ corrections in DPA has not been done yet
for $\ggwwffff$.

The combination of virtual and real-photonic corrections is non-trivial
for two reasons. First, the finite-fermion-mass effects have to be
restored in the phase-space regions of collinear photon radiation off
charged fermions, and the IR regularization for soft-photon emission
has to be implemented. To this end, we employ the dipole subtraction
formalism for photon radiation \cite{Dittmaier:1999mb,Roth:1999kk} 
as well as the
more conventional phase-space slicing approach.
The second subtlety concerns the fact that we apply the DPA only to
the virtual corrections, but not to the real-photonic parts. 
Therefore, the cancellation of soft and collinear singularities
has to be done carefully, in order to avoid mismatch.

The paper is organized as follows:
After a brief outline of our strategy in the next section,
in \refse{se:virtual} we describe the actual calculation of the virtual
corrections in DPA. Apart from the general concept, we give some
details on an efficient way for a numerically
stable evaluation, 
on renormalization issues, 
on the treatment of the $s$-channel Higgs resonance, 
and on an improved Born approximation 
used in the threshold region of W-pair production. 
Section~\ref{se:softcoll} deals with the combination
of virtual and real-photonic corrections; all relevant details for the
application of dipole subtraction and phase-space slicing to the
considered processes can be found there. Moreover, the differences
in the evaluation of collinear-safe and non-collinear-safe observables
are described. Our discussion of numerical results is presented in
\refse{se:numerics}; besides integrated cross 
sections, we also discuss angular, energy, and invariant-mass 
distributions. A summary is given in \refse{se:summary},
and the appendices provide further details on the evaluation
of coefficient functions for the factorizable virtual corrections
as well as on the generalization of the dipole subtraction method
for non-collinear-safe observables.

\section{Strategy of the calculation}
We consider the process
\beqar
\label{eq:process}
\ga(k_1,\la_1)+\ga(k_2,\la_2) &\to& \PW^+(k_+,\la_+) + \PW^-(k_-,\la_-) \nn\\
 &\to& f_1(p_1,\si_1)+\bar f_2(p_2,\si_2)+f_3(p_3,\si_3)+\bar f_4(p_4,\si_4),
\eeqar
where $k_i$ and $p_i$ denote the momenta and $\la_i$ and $\si_i$ 
the helicities of the corresponding particles.

\begin{sloppypar}
The lowest-order cross section $\rd\si_{\Born}^{\ggffff}$, 
based on the complete matrix elements $\M_{\Born}^{\ggffff}$ 
with massless fermions, has been discussed in detail in 
\citere{Bredenstein:2004ef}. Suppressing the averaging over the photon 
polarizations and the spin and colour summation for the final state
in the notation, it reads 
\beq
\int\rd\si_{\Born}^{\ggffff}=\frac{1}{2s}\int\rd\Phi_{4f}
|\M_{\Born}^{\ggffff}|^2,
\eeq
with 
\beq
s = (k_1+k_2)^2, \qquad s_{ij} = (p_i + p_j)^2, \quad i,j=1,2,3,4.
\eeq
The variables $s_{ij}$ are introduced for later use.
\end{sloppypar}

In the following we focus on the radiative corrections of $\Ord(\al)$
which consist of virtual corrections $\rd\si_{\virt}^{\ggffff}$ 
to the process \refeq{eq:process} 
and real-photonic corrections $\rd\si^{\ggffffg}$, originating from the process
\beqar
\ga(k_1,\la_1)+\ga(k_2,\la_2) &\to& \PW^+(k_+,\la_+) + \PW^-(k_-,\la_-) 
\,\,(\,+\,\ga\,)\nn\\
 &\to& f_1(p_1,\si_1)+\bar f_2(p_2,\si_2)+f_3(p_3,\si_3)+\bar f_4(p_4,\si_4)
+\ga(k,\la_\ga).
\hspace*{2em}
\label{eq:bremsprocess}
\eeqar
Combining the different contributions we obtain the $\Ord(\al)$-corrected 
prediction for the cross section,
\beq
\label{eq:cs}
\int\rd\si=\int\rd\si_{\Born}^{\ggffff} + 
\int\rd\si_{\virt}^{\ggffff} +
\int\rd\si^{\ggffffg}.
\eeq
The real-photonic corrections $\rd\si^{\ggffffg}$ are based
on the full lowest-order matrix elements $\M_{\Born}^{\ggffffg}$ 
of the process $\ggffffg$, which were calculated in 
\citere{Bredenstein:2004ef} for massless fermions. 
In the limit of vanishing photon momentum $k$ (soft limit) or when the photon 
becomes collinear to an external charged fermion (collinear limit), 
the cross section diverges. 
Considering the process $\ggffffg$ with a visible photon 
(which is neither soft nor collinear), these singularities are 
removed by imposing appropriate phase-space cuts 
which are justified by the finite experimental resolution. 
For predictions of the $\ggffffkg$ processes, i.e.\ with or without 
photon radiation, the singular phase-space regions of soft or collinear 
emission have to be integrated over. 
In this case the real corrections are combined with the virtual 
corrections which contain exactly the same singularities with opposite 
sign. The regularization of the singularities in the real corrections 
by small photon and fermion masses, $\la$ and $m_f$, as well as 
the matching with the singularities in the virtual corrections, 
is described in detail in \refse{se:softcoll}.
The starting point is a separation into a finite and a singular part,
\beq
\rd\si^{\ggffffg}=\rd\si_{\finite}^{\ggffffg}+
\rd\si_{\sing}^{\ggffffg},
\eeq
where the soft and collinear singularities appear in 
$\rd\si_{\sing}^{\ggffffg}$ as $\ln \la$ and $\ln m_f$ terms,
respectively.

The virtual corrections to the process \refeq{eq:process} are calculated 
in the DPA, which is explained in \refse{se:virtual}.
Since the real corrections are based on complete $\ggffffg$ matrix elements
(i.e.\ they are not calculated in DPA), 
the cancellation of soft and collinear singularities in Eq.~\refeq{eq:cs} 
requires particular care. To this end, we apply the DPA only 
to the finite part of the virtual corrections,
\beq
\rd\si_{\virt}^{\ggffff}\to\rd\si_{\mathrm{virt,finite,DPA}}^{\ggwwffff}+
\rd\si_{\mathrm{virt,sing}}^{\ggffff}.
\eeq
Technically this is achieved by subtracting the singular part in DPA 
from the DPA virtual corrections and adding the exact singular part 
$\rd\si_{\mathrm{virt,sing}}^{\ggffff}$. 
Of course, this procedure involves some freedom, because
finite terms can be shifted between 
$\rd\si_{\mathrm{virt,finite,DPA}}^{\ggffff}$ and 
$\rd\si_{\mathrm{virt,sing}}^{\ggffff}$. This arbitrariness is, however, 
of the order of the uncertainty $\Ord(\al\GW/(\pi\MW))$ 
of our calculation. 
In the $\Pep\Pem$ case this has been checked numerically
in \citere{Denner:2000bj}.

Inserting these rearrangements into Eq.~\refeq{eq:cs} we obtain 
\beq
\int\rd\si=\int\rd\si_{\Born}^{\ggffff} + 
\int\rd\si_{\mathrm{virt,finite,DPA}}^{\ggwwffff} +
\int\rd\si_{\mathrm{virt+real,sing}}^{\ggffff} +
\int\rd\si_{\finite}^{\ggffffg},
\eeq
where $\int\rd\si_{\mathrm{virt+real,sing}}^{\ggffff}
=\int\rd\si_{\mathrm{virt,sing}}^{\ggffff} + 
\int\rd\si_{\mathrm{real,sing}}^{\ggffff\ga}$ does not 
contain any dependence on the photon mass anymore.
Collinear singularities, appearing as $\ln m_f$ terms, also cancel
if the observable is sufficiently inclusive. 
Such {\it collinear-safe} observables result if photons within cones 
collinear to any outgoing charged fermion are treated inclusively, 
i.e.\ if they are not separated from the nearly collinear fermion by 
any phase-space or event selection cuts.
For non-collinear-safe observables 
logarithms of the fermion masses remain in the final result. This case 
demands a special treatment of the singular terms. 
We elaborate more on this issue in \refse{se:non-coll-safe}.

\begin{sloppypar}
The radiative corrections are implemented in a Monte Carlo generator
called {\sc Coffer$\gamma\gamma$}, which is based on the lowest-order
calculation described in \citere{Bredenstein:2004ef}. 
We emphasize that we have actually constructed two independent
Monte Carlo programs, each of which 
employs independent routines for the
matrix elements (with and without corrections), for the subtraction
procedure, and for the phase-space integration.
The numerical results obtained with the two programs are in mutual
agreement within statistical uncertainties.
\end{sloppypar}

\section{Virtual corrections}
\label{se:virtual}
\subsection{Concept of the double-pole approximation}
\label{se:dpaconcept}

In the DPA the matrix element for $\ggffff$ 
is expanded around the poles of the two resonant W~propagators.
The leading term of this expansion receives contributions from so-called 
factorizable and non-factorizable corrections. 
For the details of this classification, especially how a 
gauge-invariant decomposition is obtained, we refer to 
\citeres{Denner:2000bj,Aeppli:1993rs,Melnikov:1995fx,Denner:1997ia}. 

The generic Feynman diagram for the factorizable corrections is shown 
\begin{figure}
\centerline{
\begin{picture}(210,125)(0,0)
\Photon(50,80)( 25, 125){2}{4}
\Photon( 25, 35)(50, 80){2}{4}
\Photon(50,80)(170,110){2}{11}
\Photon(50,80)(170,50){2}{11}
\ArrowLine(170,110)(210, 125)
\ArrowLine(210,95)(170,110)
\ArrowLine(210,35)(170,50)
\ArrowLine(170,50)(210,65)
\GCirc(50,80){10}{.5}
\GCirc(110,95){10}{1}
\GCirc(110,65){10}{1}
\GCirc(170,110){10}{.5}
\GCirc(170,50){10}{.5}
\DashLine( 90,30)( 90,130){2}
\DashLine(130,30)(130,130){2}
\put(70,56){W}
\put(70,98){W}
\put(140,40){W}
\put(140,112){W}
\put(8,30){$\gamma$}
\put(8,125){$\gamma$}
\put(0,15){On-shell production}
\put(135,15){On-shell decays}
\end{picture} }
\caption{Generic Feynman diagram of the virtual factorizable corrections 
to $\ggwwffff$. The shaded blobs stand for loop corrections to the 
production and decay processes.}
\label{fig:fact}
\end{figure}
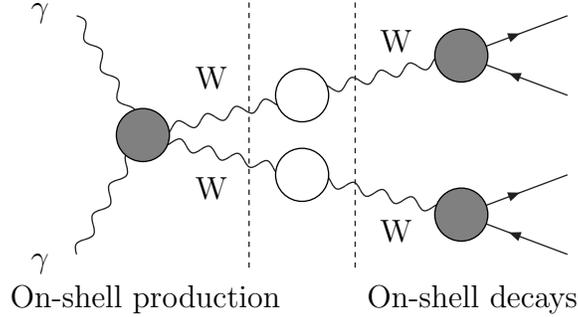
in \reffi{fig:fact}. It factorizes into the on-shell W-pair production,
the off-shell W-boson propagators, 
and the subsequent on-shell W~decays. 
The corrections can be attributed to either of these subprocesses. 
When integrating over the full $4f$ phase space, the W bosons usually 
are not on shell. 
However, a gauge-independent evaluation of the matrix elements for
production and decay requires on-shell momenta for the W~bosons.
Therefore, we have to perform an on-shell projection, 
i.e.\ the momenta of the fermions are deformed in such a way 
that the W bosons become on shell. The deformation involves a certain 
freedom and introduces an error of $\Ord(\al\GW/(\pi\MW))$. 
We define the on-shell projection by fixing the directions of the 
$\PW^+$ boson and of the fermions $f_1$ and $f_3$. The explicit formulas 
can be found in Appendix A of \citere{Denner:2000bj}.
For later use, 
we label the new momenta $\hat k_{\pm}$ and $\hat p_i$ and define 
the kinematic invariants
\beq
\hat t = (k_1-\hat k_+)^2 = (k_1-\hat p_1-\hat p_2)^2, \qquad
\hat u = 2\MW^2-s-\hat t.
\eeq

Apart from the factorizable corrections there are additional
doubly-resonant contributions.
In the corresponding diagrams subprocesses are linked by a photon. 
These diagrams become doubly resonant
in the limit of vanishing photon momentum, as can be 
seen from the soft-photon approximation in which the correction is 
proportional to the lowest-order cross section.
The relative correction factor for these
so-called non-factorizable corrections is, thus, not dependent of
the actual production mechanism of the W~pairs, but only on the
electric charges and kinematics of the external particles of the
process.
The non-factorizable corrections were calculated in 
\citeres{Melnikov:1995fx,Denner:1997ia} for $\eewwffff$. 
We can transfer the results for the $\Pep\Pem$ case by simply
omitting all contributions in which the exchanged photon is linked
to an $\Pe^\pm$ from the initial state.
The different types of relevant diagrams are depicted in \reffi{fig:nonfdiags}. 
The first two diagrams, labelled (a) and (b), are manifestly 
non-factorizable, i.e.\ the photon links different subprocesses 
so that the propagators in the diagrams cannot be factorized anymore. 
The diagrams (c), (d), and (e) contain both factorizable and
non-factorizable contributions. Their factorizable parts are
defined as the residues for on-shell W~bosons times the off-shell
W-boson propagators; note that this procedure introduces artificial
soft IR divergences connected with the on-shellness of the W~bosons
in the loops. The non-factorizable parts of the diagrams are obtained
from the difference of
the doubly-resonant contribution of the full diagrams and their
factorizable parts; the artificially introduced IR divergences of the
factorizable parts are, thus, compensated by corresponding terms
in the non-factorizable parts.

Following this strategy, the virtual corrections in DPA can be written as
\beqar
\label{eq:virtual}
\rd\si_{\mathrm{virt,DPA}}^{\ggwwffff}
&=& \frac{1}{2s}\int\rd\Phi_{4f}\biggl(
2\Re\{\delta\M_{\mathrm{virt,fact}}\M_{\Born,\DPA}^*\} \nn\\
&& \qquad\qquad\quad {} + \delta_{\mathrm{virt,nfact}}|\M_{\Born,\DPA}|^2
+ |\delta\M_{\mathrm{Higgs}}|^2\biggr),
\eeqar
where $\M_{\Born,\DPA}$ denotes the tree-level matrix element in DPA 
and $\delta_{\mathrm{virt,nfact}}$ contains the non-factorizable corrections. 
The factorizable corrections $\delta\M_{\mathrm{virt,fact}}$ also contain 
a contribution of the $s$-channel Higgs resonance, $\delta\M_{\mathrm{Higgs}}$. 
In order to describe this resonance properly, it is not sufficient to 
include the interference of $\delta\M_{\mathrm{Higgs}}$ with the Born
matrix element, but the square of this matrix-element contribution
has to be taken into account in addition. To this end, 
$\delta\M_{\mathrm{Higgs}}$ has to be defined in a gauge-invariant way.
Our treatment of $\delta\M_{\mathrm{Higgs}}$ is described
in \refse{se:higgs} in detail.

\subsection{Factorizable corrections}

\subsubsection{Calculation of the one-loop amplitudes} 
\label{se:oneloopcalc}

The factorizable corrections comprise the corrections to the on-shell 
production of the W bosons and their on-shell decay and can be 
expressed as
\beqar
\label{eq:fact}
\de\M_{\mathrm{virt,fact}}&=&
\sum_{\la_+,\la_-}\frac{1}{K_+K_-}\left(
\de\M^{\ggww}\M_{\Born}^{\PW^+\to f_1\bar f_2}
\M_{\Born}^{\PW^-\to f_3\bar f_4} \right. 
\nn\\
&& \qquad\qquad\qquad {} + \M_{\Born}^{\ggww}\de\M^{\PW^+\to f_1\bar f_2}
\M_{\Born}^{\PW^-\to f_3\bar f_4} 
\nn\\
&& \qquad\qquad\qquad {} + \left. \M_{\Born}^{\ggww}\M_{\Born}
^{\PW^+\to f_1\bar f_2}\de\M^{\PW^-\to f_3\bar f_4}\right),
\eeqar
where we introduced the abbreviations 
\beq
K_{\pm}=k_{\pm}^2-\MW^2+\ri\MW\GW,
\eeq
and $\delta\M$ denote one-loop matrix elements.
Note that all matrix elements on the r.h.s.\ of Eq.~\refeq{eq:fact}
depend on the on-shell projected 
momenta, but the momenta in $K_{\pm}$ remain unchanged.
The results for the different one-loop corrections are already 
known in the literature 
\cite{Denner:1994ca,Denner:1995jv,Jikia:1996uu,Bardin:1986fi}. 
Combining them in Eq.~\refeq{eq:fact} is, however, 
non-trivial since the polarizations of 
the W~bosons have to be defined consistently in a common reference frame.

The one-loop corrections $\de\M^{\PW\to f_i\bar f_j}$ to the W~decays
are rather simple. In the massless limit they are proportional to the
respective Born matrix elements $\M_{\Born}^{\PW\to f_i\bar f_j}$,
\beq
\de\M^{\PW\to f_i\bar f_j}(\la_\PW,\hat p_i,\hat p_j) = 
\delta^{\PW\to f_i\bar f_j} \,
\M_{\Born}^{\PW\to f_i\bar f_j}(\la_\PW,\hat p_i,\hat p_j),
\eeq
where $\delta^{\PW\to f_i\bar f_j}$ is a constant correction factor that
neither depends on the kinematics nor on the helicity $\la_\PW$ of
the decaying W~boson.

The one-loop correction $\de\M^{\ggww}$ to the W-pair production process 
contains the complicated part; we have derived it in two independent ways. 
One calculation is based on the results of \citere{Denner:1995jv} which 
were obtained in a gauge with a non-linear gauge fixing term. 
The other is a new calculation based on the program 
{\sl FeynArts} \cite{FA} for the generation of the amplitudes
and on in-house {\sl Mathematica} routines for their algebraic reduction. 
This second calculation
has been carried out in the `t~Hooft--Feynman gauge and repeated
in the background-field gauge \cite{Denner:1994xt} to get an additional
consistency check.
The results obtained from the
different calculations are in mutual numerical agreement.

\begin{sloppypar}
In the following we describe an efficient way for calculating the
contribution of $\de\M^{\ggww}$ to $\de\M_{\mathrm{virt,fact}}$
of Eq.~\refeq{eq:fact}, taking into account all spin correlations.
As described in \citere{Denner:1995jv}, the matrix element $\de\M^{\ggww}$ 
for on-shell W-pair production is decomposed into a sum of
products of form factors $F_j$,
which only depend on the kinematic variables $s$ and $\hat t$,
and a set of standard matrix elements (SME) $\M_j^{\ggww}$, which contain the 
polarizations and momenta of the external photons and W~bosons,
\beqar
\label{eq:decomp}
\lefteqn{\delta \M^{\ggww}(k_1,k_2,\la_1,\la_2;\hat k_+,\hat k_-,\la_+,\la_-)}
&&
\nn\\*
&=& \sum_{j=1}^{36} F_j(s,\hat t)
\M_j^{\ggww}(k_1,k_2,\la_1,\la_2;\hat k_+,\hat k_-,\la_+,\la_-).
\eeqar
The SME $\M_j^{\ggww}$ 
are obtained from the 83 basic matrix 
elements given in Section~2 of \citere{Denner:1995jv} which are 
reduced to 36 matrix elements as described there%
\footnote{The on-shell momenta $\hat k_\pm$ and the helicities
$\la_\pm$ of the W~bosons
are denoted $k_{3,4}$ and $\la_{3,4}$ in \citere{Denner:1995jv}.}.
The decay matrix elements $\M_{\Born}^{\PW\to f_i\bar f_j}$, which
multiply $\delta\M^{\ggww}$ in Eq.~\refeq{eq:fact}, can be
included by replacing the W~polarization vectors $\veps^*_\pm$
in the definitions of the SME $\M_j$ by the ``effective 
polarization vectors''
\beq
\hat\veps^{*\,\mu}_+=\frac{e}{\sqrt{2}\sw}\frac{1}{K_+}\bar u(\hat p_1)\ga^{\mu}\om_-
v(\hat p_2), \qquad
\hat\veps^{*\,\mu}_-=\frac{e}{\sqrt{2}\sw}\frac{1}{K_-}\bar u(\hat p_3)\ga^{\mu}\om_-
v(\hat p_4),
\eeq
where $\bar u(\hat p_i)$ and $v(\hat p_i)$ are the Dirac spinors 
of the fermions and antifermions
and $\om_-=\frac{1}{2}(1-\ga_5)$ is the left-handed chirality projector.
The effective W-polarization vectors $\hat\veps^*_\pm$ are formal shorthands
for the W~propagators and the tree-level decay matrix elements, which
involve the usual SU(2) gauge coupling $e/\sw$.
Upon substituting $\veps^*_\pm\to\hat\veps^*_\pm$ in the SME for on-shell
W-pair production,
we obtain a new set of SME $\M_j$ that correctly transfer
the W~polarization to the decay,
\beqar
\M_j(k_1,k_2,\la_1,\la_2;k_+^2,k_-^2;\{\hat p_i\}) &=&
\M_j^{\ggww}(k_1,k_2,\la_1,\la_2;\hat k_+,\hat k_-,\la_+,\la_-)
\Big|_{\veps^*_\pm\to\hat\veps^*_\pm}
\nn\\
&=& \sum_{\la_+,\la_-}\frac{1}{K_+K_-}\,
\M^{\ggww}_j(k_1,k_2,\la_1,\la_2;\hat k_+,\hat k_-,\la_+,\la_-)
\nn\\
&& \qquad {} \times
\M_{\Born}^{\PW^+\to f_1\bar f_2}(\la_+,\hat p_1,\hat p_2)\,
\M_{\Born}^{\PW^-\to f_3\bar f_4}(\la_-,\hat p_3,\hat p_4).
\nn\\
\eeqar
The new SME $\M_j$ can be easily evaluated with spinor methods, 
as e.g.\ described in \citere{Dittmaier:1998nn}.
\end{sloppypar}

In summary the factorizable part of the virtual correction takes the
form
\beqar
\label{eq:fact2}
\delta \M_{\mathrm{virt,fact}}
&=& \sum_{j=1}^{36} F_j(s,\hat t)
\M_j(k_1,k_2,\la_1,\la_2;k_+^2,k_-^2;\{\hat p_i\})
\nn\\
&& {}
+ \left( \delta^{\PW^+\to f_1\bar f_2}+ \delta^{\PW^-\to f_3\bar f_4} \right)
\M_{\Born,\DPA}(k_1,k_2,\la_1,\la_2;k_+^2,k_-^2;\{\hat p_i\}).
\eeqar

\subsubsection{Details of the numerical evaluation}
\label{se:virtualeval}

The formulas for the coefficient functions $F_j$ are 
rather lengthy and contain many one-loop integrals, which in turn
involve many dilogarithmic functions, etc. Thus, to speed up the
numerical evaluation it is desirable not to evaluate the $F_j$
at each phase-space point.
Moreover, numerical instabilities occur at the boundary of the 
phase space where the scattering angle $\theta$
between the W bosons and the beam axis tends to 0 or $\pi$. This is 
due to the inverse Gram determinants appearing in the 
Passarino--Veltman reduction \cite{Passarino:1978jh} 
of the tensor integrals. The problems of speed and stability
can be solved by expanding the functions $F_j(s,\hat t)$ in terms
of a generalized Fourier series
in the variable $\hat t$ for fixed values of $s$. 
The coefficients of this expansion are calculated before the 
Monte Carlo integration. An appropriate system of orthogonal
functions in the variable $x=\cos\theta$, which is equivalent to
a function of $\hat t$ for fixed $s$, is provided by the Legendre 
polynomials
\beq
P_l(x)=\frac{1}{2^ll!}\frac{\rd^l}{\rd x^l}\left[(x^2-1)^l\right],
\qquad l=0,1,...\quad .
\eeq
For this basis functions, the coefficients read
\beq
\label{eq:legcoeff}
c_{j,l}(s)=\frac{2l+1}{2}\int_{-1}^{+1}\rd\cos{\theta}\,
(\hat t-\MW^2)(\hat u-\MW^2) F_j(s,\hat t)P_l(\cos{\theta}),
\eeq
where we have introduced the factor $(\hat t-\MW^2)(\hat u-\MW^2)$
in order to flatten the $t$- and $u$-channel poles in the functions $F_j$.
This improves the efficiency of the expansion.
The integration in Eq.~\refeq{eq:legcoeff} is carried out using Gaussian 
integration. 
With 40 integration points the region of instability is not entered 
(for energies up to a few TeV), and 
the integration is sufficiently precise. 
During the Monte Carlo integration the coefficient functions are 
recovered by the generalized Fourier series
\beq
F_j(s,\hat t)=\sum_{l=0}^{\infty}\frac{1}{(\hat t-\MW^2)(\hat u-\MW^2)}\,
c_{j,l}(s) P_l(\cos{\theta}).
\eeq

In \citere{Denner:2000bj} the same concept was used to evaluate
the factorizable corrections to $\eewwffff$; 
there it was sufficient to use the Legendre 
polynomials up to $l=20$ for a good accuracy. 
In the case of 
$\ggtoww$, however, the coefficient functions involve inverse Gram 
determinants $1/(\hat t\hat u-\MW^4)\propto1/\sin^2\theta$ 
which appear in the Passarino--Veltman reduction of the tensor integrals. 
As each step in this recursive reduction involves such an inverse 
determinant, $1/(\hat t\hat u-\MW^4)$ can appear up to the fourth power. 
At $\cos{\theta}\approx\pm1$ this factor leads to a behaviour of the 
$F_j(s,\hat t)$ that is not well approximated by the Legendre expansion.
Using higher-order Legendre polynomials is not a solution since this 
increases the calculation time and also requires more integration 
points for the Gaussian integration. The more points are used in the Gaussian  
integration, the closer some of these points approach the integration boundary 
where the numerical stability of the coefficient function breaks down. 
Therefore, we follow a different strategy based on the fact that 
the helicity amplitudes for the on-shell process $\ggtoww$ are
smooth functions of $\cos{\theta}$, apart from the $t$- and $u$-channel
poles.
Thus, within the full amplitude the factors 
$1/(\hat t\hat u-\MW^4)$ have to cancel between contributions of different 
coefficient functions. 
To make use of this fact we change the basis of SME by a linear 
transformation in such a way that the new coefficient functions 
correspond to helicity amplitudes of the on-shell process
$\ggtoww$. Some details of this transformation can be found in 
\refapp{app:sme}.
After this transformation the uncertainty of the approximated matrix elements 
in Eq.~\refeq{eq:decomp} is well below $10^{-4}$ with respect to the 
Born matrix elements for all values of $\cos{\theta}$. 

In contrast to the $\Pep\Pem$ case, the CM energy 
$\sqrt{s}$ of the photons is not fixed. Thus, we have to perform the
Legendre expansions for different values of $s$. During the 
Monte Carlo integration we derive an approximate value of 
the coefficients $c_{j,l}(s)$ by interpolation. 
Since the $F_j(s,\hat t)$ depend on $s$ very smoothly, it is sufficient to 
calculate the $c_{j,l}(s)$ at intervals of $\Delta s\lsim 1\GeV$. 
In these intervals we then interpolate with a polynomial of third order. 
We have checked that, up to $1\TeV$, this yields a sufficient accuracy 
(i.e.\ better than the accuracy of the Legendre expansion).

\subsubsection{Renormalization and imaginary parts of virtual corrections}

For on-shell W-pair production, which was considered in 
\citere{Denner:1995jv}, imaginary parts of counterterms, if included,
do not influence the correction to the matrix element square.
The reason is that
for the $2\to2$ scattering process $\ggtoww$ all SME, and thus also
the Born matrix element, can be taken real by appropriate phase choices.
Thus, the operation of taking the real part in the interference term
$2\Re\{\M_{\mathrm{ct}}\M_{\Born}^*\}$ of the counterterm contribution
$\M_{\mathrm{ct}}$ to the one-loop amplitude with the Born amplitude
effectively acts on the
renormalization constants themselves. The same argument shows that also
imaginary parts of loop integrals drop out. 
These arguments are no longer true if the decay of the W~bosons is taken
into account, because the SME and the Born matrix element
$\M_{\Born,\DPA}$ become necessarily complex. Thus, imaginary parts
of renormalization constants and of loop integrals in general matter.
Considering the W-decay amplitudes in the DPA in more detail, as 
e.g.\ done in \citere{Beenakker:1998gr} for the $\Pep\Pem$ case,
one can see that imaginary parts average to zero after the
azimuthal decay angles of the W-decay products are integrated over.

We have calculated the virtual corrections taking into 
account the imaginary parts of all loop integrals. 
As already mentioned, we carried out the whole loop calculation in
different gauges: in the `t~Hooft--Feynman gauge \cite{Denner:1993kt}, 
in a non-linear gauge \cite{Denner:1995jv}, and in the background-field 
gauge \cite{Denner:1994xt}. We find agreement between the results
obtained in these different gauges, but only if we also take into account
the imaginary parts of the loops that contribute to renormalization
constants. In order to explain this fact, we consider the
counterterm contributions to the one-loop matrix element in more detail.

Following \citere{Denner:1995jv}, 
we write the Born matrix element in DPA as
\beq
\M_{\Born,\DPA}=8\pi\al\left\{\frac{s}{\MW^2-\hat t}\M_{0,t}
+\frac{s}{\MW^2-\hat u}\M_{0,u}
-(\veps_1\veps_2)(\hat\veps^*_+\hat\veps^*_-)\right\},
\eeq
where $\M_{0,t}$ and $\M_{0,u}$ are abbreviations for specific
combinations of momenta and polarization vectors defined as in Eq.~(22) of
\citere{Denner:1995jv} for on-shell W-pair production.
In the `t~Hooft--Feynman gauge, the counterterm contribution to the
production part of the factorizable correction reads
\beqar
\de\M_{\mathrm{ct,prod}}^{\mathrm{tHF}} &=& 
\M_{\Born,\DPA}\left(2\de Z_e+\de Z_W+\de Z_{AA}
-\frac{\cw}{\sw}\de Z_{ZA}\right) \nn\\
&& {} -8\pi\al\left(\frac{s\de\MW^2}{(\hat t-\MW^2)^2}\M_{0,t}+
\frac{s\de\MW^2}{(\hat u-\MW^2)^2}\M_{0,u}\right) \nn\\
&& {} -4\pi\al\left(\frac{(\veps_1\hat\veps^*_+)(\veps_2\hat\veps^*_-)}
{(\hat t-\MW^2)}+\frac{(\veps_1\hat\veps^*_-)(\veps_2\hat\veps^*_+)}
{(\hat u-\MW^2)}\right)
\left(2\de\MW^2+\frac{\MW^2}{\sw\cw}\de Z_{ZA}\right) \nn\\
&& {} +4\pi\al\frac{e\MW}{2\sw}
\left(\frac{(\veps_1\hat\veps^*_+)(\veps_2\hat\veps^*_-)}{(\hat t-\MW^2)^2}
+\frac{(\veps_1\hat\veps^*_-)(\veps_2\hat\veps^*_+)}
{(\hat u-\MW^2)^2}\right)\de t,
\label{eq:ctthf}
\eeqar
where we adopt the conventions of \citere{Denner:1993kt} 
for the renormalization constants $\de Z_e$, $\de Z_W$, etc.
The explicit
calculation of the constants in terms of self-energies is also
described there.
The counterterm contribution in the background-field gauge \cite{Denner:1994xt}
can be obtained
from $\de\M_{\mathrm{ct,prod}}^{\mathrm{tHF}}$ by simply omitting the
$\de Z_{ZA}$ terms, because $\de Z_{ZA}$ vanishes owing to the
background-field gauge invariance.
In the non-linear gauge the counterterm contribution reads
\beqar
\de\M_{\mathrm{ct,prod}}^{\mathrm{NL}} &=& 
\M_{\Born,\DPA}\left(2\de Z_e+\de Z_W+\de Z_{AA}
-\frac{\cw}{\sw}\de Z_{ZA}\right) \nn\\
&& {} -8\pi\al\left(\frac{s\de\MW^2}{(\hat t-\MW^2)^2}\M_{0,t}
+\frac{s\de\MW^2}{(\hat u-\MW^2)^2}\M_{0,u}\right),
\label{eq:ctnl}
\eeqar
as described in \citere{Denner:1995jv}, which is different from its
counterpart in `t~Hooft--Feynman gauge. Note also that the explicit
expressions of the renormalization constants in the different gauges
are in general different.

Imaginary parts of loop and counterterm contributions that are
proportional to the Born matrix element, $\de\M=c\M_{\Born}$,
cannot influence matrix element squares, because
$2\Re\{\de\M\M_{\Born}^*\}=2\Re\{c\}|\M_{\Born}|^2$.
Thus, the W-mass renormalization constant $\de\MW^2$ is the only 
renormalization constant whose imaginary part plays a role, since
the tadpole counterterm $\de t$ is a real quantity.
From Eqs.~\refeq{eq:ctthf} and \refeq{eq:ctnl}, we see that 
$\de\MW^2$, which is equal in all three considered gauges, enters the
counterterm contributions in the `t~Hooft--Feynman gauge and 
in the non-linear gauge in different ways. In fact, we have
checked numerically that the virtual corrections in these two
gauges are different (though finite) if the usual on-shell prescription
$\de\MW^2=\Re\{\Sigma^W_{\mathrm{T}}(\MW^2)\}$ 
(see e.g.\ \citere{Denner:1993kt}) is applied, where
$\Sigma^W_{\mathrm{T}}(k^2)$ is the transverse part of the W-boson
self-energy with momentum transfer $k$.
If we, on the other hand, use the definition
$\de\MW^2=\Sigma^W_{\mathrm{T}}(\MW^2)$, i.e.\ without taking the
real part of the self-energy, we find agreement for the results 
from the different gauges. This clearly shows that the imaginary
part of a one-loop amplitude is in general gauge dependent if
imaginary parts in renormalization constants are not taken into
account. The reason for this fact, in other words, is that the
decomposition of a renormalized transition matrix element into
genuine loop parts and counterterm contributions depends on the
gauge fixing.%
\footnote{A consistent renormalization prescription with complex 
renormalization constants naturally leads to complex masses for
unstable particles. Such a renormalization scheme was proposed
recently in \citere{Denner:2005fg} in the context of a full ${\cal O}(\alpha)$
calculation for $\Pep\Pem\to4f$.}

In our Monte Carlo generator we have taken into account the imaginary parts
of the virtual corrections (including the ones from counterterms);
more precisely they can be switched on and off optionally.
As explained above, they could only affect observables that are sensitive
to the azimuthal decay angles of the fermions. In our numerical results,
we could, however, find no significant effects.

\subsubsection{Higgs resonance}
\label{se:higgs}

The loop-induced Higgs resonance, $\ga\ga\to\PH\to\PW\PW\to 4f$, 
belongs to the class of factorizable contributions. 
Nevertheless, its treatment, 
especially the question of gauge invariance when including the Higgs 
decay width, deserves some care. 
In \citere{Denner:1995jv} the diagrams with an $s$-channel 
Higgs resonance were decomposed into a gauge-invariant resonant part 
and a gauge-dependent non-resonant part.
If we write the contribution of the Higgs-exchange diagrams as 
\beq
\delta\M^{\ga\ga\PH}=\frac{F^H(s)}{s-\MH^2}
(\veps_1\veps_2)(\hat\veps^*_+\hat\veps^*_-),
\eeq
with $F^H(s)$ given in Section 4.3 of \citere{Denner:1995jv}, and 
$\veps_1$ and $\veps_2$ being the polarization vectors of the photons, 
then the Higgs decay width can be introduced by replacing 
\beq
\label{eq:higgs}
\delta\M^{\ga\ga\PH}\to\left(\frac{F^H(\MH^2)}{s-\MH^2+\ri\MH\GH}
+\frac{F^H(s)-F^H(\MH^2)}{s-\MH^2}\right)(\veps_1\veps_2)
(\hat\veps^*_+\hat\veps^*_-).
\eeq
As the residue $F^H(\MH^2)$ is gauge independent, we have introduced 
the Higgs decay width $\GH$ in a gauge-invariant way. 
Recall that the choice of the polarization vectors of the photons is 
such that they obey 
\beq
\veps_ik_j=0, \qquad i,j=1,2.
\eeq

Close to the resonance, the contribution of the Higgs-exchange diagrams 
is strongly enhanced. This is why we also take into account the 
square of the resonant part in Eq.~\refeq{eq:virtual},
\beq
\label{eq:higgs2}
\delta\M_{\mathrm{Higgs}}=\frac{F^H(\MH^2)(\veps_1\veps_2)
(\hat\veps^*_+\hat\veps^*_-)}
{s-\MH^2+\ri\MH\GH}.
\eeq

In this approach only the leading contribution to the Higgs resonance
is taken into account. 
However, the gauge-invariant separation of $\delta\M_{\mathrm{Higgs}}$ 
from the remaining one-loop amplitude easily allows for specific improvements
in predictions for the Higgs-production signal in the future.
To this end, a pole expansion about the Higgs resonance would be an adequate
first step. Conceptually this expansion again leads to factorizable and 
non-factorizable contributions, but the corresponding ingredients are not
all available yet and their calculation is beyond the scope of this work.
It should be mentioned that both the ${\cal O}(\alpha)$ electroweak and 
${\cal O}(\alpha_{\mathrm{s}})$ QCD 
virtual factorizable corrections to (on-shell) Higgs production
$\gamma\gamma\to\PH$ can be deduced from the corresponding two-loop
calculations \cite{Djouadi:1990aj}
(see also references therein)
for the decay $\PH\to\gamma\gamma$.

\subsection{Non-factorizable corrections}

As explained in \refse{se:dpaconcept}, 
we make use of the result for the non-factorizable 
corrections to $\eewwffff$.
According to \citeres{Denner:2000bj,Denner:1997ia} we write 
the correction factor to the lowest-order cross section as 
a sum over contributions that are associated with 
different pairs of fermions,
\beq
\delta_{\mathrm{virt,nfact}}=\sum_{a=1,2}\sum_{b=3,4}(-1)^{a+b+1}
Q_aQ_b\frac{\al}{\pi}\Re\left\{\Delta^{\virt}(k_+,p_a;k_-,p_b)\right\}.
\eeq
\bfi
\begin{center}
\begin{picture}(360,380)(0,10)
\Text(0,365)[lb]{(a) type (\mfp)}
\put(20,260){
\begin{picture}(150,100)(0,0)
\Photon(30,50)( 5, 95){2}{5}
\Photon(30, 50)( 5, 5){-2}{5}
\Photon(30,50)(90,20){2}{6}
\Photon(30,50)(90,80){-2}{6}
\Vertex(60,65){2.0}
\GCirc(30,50){10}{.5}
\Vertex(90,80){2.0}
\Vertex(90,20){2.0}
\ArrowLine(90,80)(120, 95)
\ArrowLine(120,65)(90,80)
\ArrowLine(120, 5)( 90,20)
\ArrowLine( 90,20)(105,27.5)
\ArrowLine(105,27.5)(120,35)
\Vertex(105,27.5){2.0}
\Photon(60,65)(105,27.5){-2}{5}
\put(86,50){$\gamma$}
\put(63,78){$\PW$}
\put(40,65){$\PW$}
\put(52,18){$\PW$}
\put(12, 5){$\ga(k_1)$}
\put(12,90){$\ga(k_2)$}
\put(125,90){$f_1(p_1)$}
\put(125,65){$\bar f_2(p_2)$}
\put(125,30){$f_3(p_3)$}
\put(125, 5){$\bar f_4(p_4)$}
\end{picture}
}
\Text(210,365)[lb]{(b) type (\ffp)}
\put(230,260){
\begin{picture}(120,100)(0,0)
\Photon(30,50)( 5, 95){2}{5}
\Photon(30, 50)( 5, 5){-2}{5}
\Photon(30,50)(90,80){-2}{6}
\Photon(30,50)(90,20){2}{6}
\GCirc(30,50){10}{.5}
\Vertex(90,80){2.0}
\Vertex(90,20){2.0}
\ArrowLine(90,80)(120, 95)
\ArrowLine(120,65)(105,72.5)
\ArrowLine(105,72.5)(90,80)
\Vertex(105,72.5){2.0}
\ArrowLine(120, 5)( 90,20)
\ArrowLine( 90,20)(105,27.5)
\ArrowLine(105,27.5)(120,35)
\Vertex(105,27.5){2.0}
\Photon(105,27.5)(105,72.5){2}{4.5}
\put(93,47){$\gamma$}
\put(55,73){$\PW$}
\put(55,16){$\PW$}
\end{picture}
}
\Text(0,235)[lb]{(c) type (\mmp)}
\put(20,130){
\begin{picture}(120,100)(0,0)
\Photon(30,50)( 5, 95){2}{5}
\Photon(30, 50)( 5, 5){-2}{5}
\Photon(30,50)(90,80){-2}{6}
\Photon(30,50)(90,20){2}{6}
\Photon(70,30)(70,70){2}{3.5}
\Vertex(70,30){2.0}
\Vertex(70,70){2.0}
\GCirc(30,50){10}{.5}
\Vertex(90,80){2.0}
\Vertex(90,20){2.0}
\ArrowLine(90,80)(120, 95)
\ArrowLine(120,65)(90,80)
\ArrowLine(120, 5)( 90,20)
\ArrowLine( 90,20)(120,35)
\put(76,47){$\gamma$}
\put(45,68){$\PW$}
\put(45,22){$\PW$}
\put(72,83){$\PW$}
\put(72,11){$\PW$}
\end{picture}
}
\Text(210,235)[lb]{(d) type (mf)}
\put(230,130){
\begin{picture}(120,100)(0,0)
\Photon(30,50)( 5, 95){2}{5}
\Photon(30, 50)( 5, 5){-2}{5}
\Photon(30,50)(90,80){-2}{6}
\Photon(30,50)(90,20){2}{6}
\Vertex(70,70){2.0}
\GCirc(30,50){10}{.5}
\Vertex(90,80){2.0}
\Vertex(90,20){2.0}
\ArrowLine(90,80)(105,87.5)
\Vertex(105,87.5){2.0}
\ArrowLine(105,87.5)(120, 95)
\ArrowLine(120,65)(90,80)
\ArrowLine(120, 5)(90,20)
\ArrowLine(90,20)(120,35)
\PhotonArc(87.5,78.75)(19.566,26.565,206.565){2}{6}
\put(57,86){$\gamma$}
\put(77,62){$\PW$}
\put(50,48){$\PW$}
\put(55,16){$\PW$}
\end{picture}
}
\Text(0,105)[lb]{(e) type (mm)}
\put(20,0){
\begin{picture}(120,100)(0,0)
\Photon(30,50)( 5, 95){2}{5}
\Photon(30, 50)( 5, 5){-2}{5}
\Photon(30,50)(90,80){-2}{6}
\Photon(30,50)(90,20){2}{6}
\Vertex(75,72.5){2.0}
\Vertex(50,60){2.0}
\GCirc(30,50){10}{.5}
\Vertex(90,80){2.0}
\Vertex(90,20){2.0}
\ArrowLine(90,80)(120, 95)
\ArrowLine(120,65)(90,80)
\ArrowLine(120, 5)( 90,20)
\ArrowLine( 90,20)(120,35)
\PhotonArc(62.5,66.25)(13.975,26.565,206.565){-2}{3.5}
\put(55,90){$\gamma$}
\put(44,45){$\PW$}
\put(62,54){$\PW$}
\put(82,64){$\PW$}
\put(55,16){$\PW$}
\end{picture}
}
\end{picture}
\end{center}
\caption{A representative set of diagrams contributing to the 
virtual non-factorizable corrections. The shaded blobs stand for all
tree-level structures contributing to $\ggtoww$.}
\label{fig:nonfdiags}
\efi
The function $\Delta^{\virt}$ receives contributions from the different types 
of diagrams in \reffi{fig:nonfdiags},
\beq
\Delta^{\virt}=\Delta^{\virt}_{\mfp}+\Delta^{\virt}_{\ffp}
+\Delta^{\virt}_{\mmp}+\Delta^{\virt}_{\mathrm{mf}}
+\Delta^{\virt}_{\mathrm{mm}},
\eeq
for which the results were given in terms of scalar integrals 
in \citere{Denner:2000bj}. The final result for $a=2,b=3$ 
(all other contributions can be derived by appropriate substitutions) is 
\beqar
\lefteqn{\Delta^{\virt}_{\mfp}+\Delta^{\virt}_{\ffp}
+\Delta^{\virt}_{\mathrm{mf}}}\qquad \nn\\
&\sim& {} -\frac{K_+K_-s_{23}\det(Y_0)}{\det(Y)}
D_0(-p_4,k_++p_3,p_2+p_3,0,M,M,0)\nn\\
&& {}-\frac{K_+\det(Y_3)}{\det(Y)}F_3-\frac{K_-\det(Y_2)}{\det(Y)}F_2
+\ln\left(\frac{\la^2}{\MW^2}\right)\ln\left(-\frac{s_{23}}{\MW^2}
-\ri\eps\right),
\nn\\
\Delta_{\mathrm{mm'}}^{\virt}
&\sim& (2\MW^2-s)\left\{ C_0(k_+,-k_-,0,M,M)
		-C_0(k_+,-k_-,\la,\MW,\MW)\Big|_{k_\pm^2=\MW^2}\right\},
\nn\\
\Delta_{\mathrm{mm}}^{\virt}
&\sim& 2\ln\left(\frac{\la\MW}{-K_+}\right)+2\ln\left(\frac{\la\MW}
{-K_-}\right)+4,
\label{eq:Deltanf}
\eeqar
where the sign ``$\sim$'' indicates that the limit $k_{\pm}^2\to\MW^2$ 
and $\GW\to 0$ is carried out whenever this does not lead to a singularity.
The matrices $Y_0$, $Y_2$, $Y_3$, and $Y$ arise from the reduction of 
5-point functions and can be found in Section 3.1 of 
\citere{Denner:1997ia}. The functions $F_2$ and $F_3$ are defined in 
Section 4.2, and the $C_0$ and $D_0$ functions in Appendix C.1
of the same reference. 
The contribution $\Delta^{\virt}_{\mmp}$ contains the
difference of the full off-shell and on-shell
Coulomb singularity, as described there in detail.

\begin{sloppypar}
The full correction factor $\delta_{\mathrm{virt,nfact}}$ does
not contain fermion-mass singularities \cite{Denner:2000bj},
but involves IR-singular terms $\ln\la$, as explicitly visible
in Eq.~\refeq{eq:Deltanf}. 
The latter originate from the subtraction of the virtual factorizable
correction, which involves the one-loop matrix elements for $\ggtoww$
and $\PW\to f\bar f'$
with on-shell W~bosons, from the doubly-resonant part of the
matrix element for the full $\ggffff$ process. Specifically,
the $\ln\la$ terms stem from diagrams with photon exchange between
an on-shell W~boson and another on-shell particle.
As already explained in \refse{se:dpaconcept}, these singularities
cancel in the sum of factorizable and non-factorizable contributions,
since they are artificially introduced in the corresponding
decomposition of the virtual correction.
\end{sloppypar}

\subsection{Leading universal corrections and input-parameter scheme}
\label{se:universal}

We parametrize the cross section in such a way that
the universal corrections arising from the running of the electromagnetic
coupling $\alpha$ 
and from the $\rho$-parameter are absorbed in the lowest order.
To this end, we take all particle masses as input, from which the
weak mixing angle is derived via the on-shell condition
\beq
\sw^2=1-\cw^2=1-\frac{\MW^2}{\MZ^2}.
\eeq
The electromagnetic coupling $\alpha$ is chosen in order to absorb
some universal corrections.

As pointed out in \citere{Denner:1995jv}, the relevant coupling for the
$\ggtoww$ production process is the fine-structure constant $\alpha(0)$,
because the external on-shell photons do not induce any running in their
coupling to the W~bosons.
This means that the on-shell renormalization is carried out
precisely as described in \citeres{Denner:1994xt,Denner:1993kt}
for this contribution.

For the decay of the W~bosons, it is, however, appropriate to derive
$\alpha$ from the Fermi constant $\GF$ leading to
\beq
\alpha_{\GF}=\frac{\sqrt{2}\GF\MW^2\sw^2}{\pi}.
\eeq
This modification of the coupling induces an additional finite
contribution to the charge renormalization constant,
\beq
\delta Z_{\Pe}|_{\GF}=\delta Z_{\Pe}|_{\al(0)}-\frac{1}{2}\Delta r,
\eeq
where $\delta Z_{\Pe}|_{\al(0)}$ is the charge renormalization constant
of the on-shell renormalization schemes \cite{Denner:1994xt,Denner:1993kt}
with $\alpha(0)$ as renormalized coupling.
The quantity $\Delta r$ contains the radiative corrections to muon decay;
explicit expressions for $\Delta r$ can, e.g., be found in
\citeres{Denner:1993kt,Sirlin:1980nh}. 

In summary, our lowest-order cross section scales like 
$\al(0)^2\al_{\GF}^2$.
For the relative $\Oa$ corrections we use $\al(0)$, which is the correct
effective coupling for real photon emission, so that the corrected
cross section scales like $\al(0)^3\al_{\GF}^2$.
For the loop-induced Higgs resonance we exceptionally take
the scaling factor $\al(0)^2\al_{\GF}^3$, which accounts for the
two ``photonic'' and the three ``weak'' couplings in the corresponding diagrams.
We perform this rescaling, of course, only in the gauge-invariant 
resonant part $\delta\M_{\mathrm{Higgs}}$ of the one-loop amplitude,
as defined in Eq.~\refeq{eq:higgs2}.

\subsection{Improved Born approximation}
\label{se:IBA}

\begin{sloppypar}
The motivation for calculating the virtual corrections in DPA 
lies in the dominance of doubly-resonant diagrams. 
At threshold, however, singly-resonant 
and non-resonant diagrams become equally important, thus, rendering 
the naive error estimate of $\Ord(\al\GW/(\MW\pi))$ unreliable. 
As a consequence, we decided to use the DPA only for a CM energy 
$\sqrt{s_{\ga\ga}}>170\GeV$ when integrating over the photon spectrum.
For $\sqrt{s_{\ga\ga}}<170\GeV$ we make use of an improved Born 
approximation (IBA), i.e.\ we approximate the $\Ord(\al)$ corrections 
by universal contributions without any expansion about the W resonances. 
Assuming that the IBA accounts for all $\Ord(\al)$ corrections with
pronounced enhancement factors, the relative uncertainty of the IBA 
is about $\sim\pm 2\%$. For the corresponding $\Pep\Pem$ reaction this
expectation was confirmed by the full $\Ord(\al)$ calculation
\cite{Denner:2005es,Denner:2005fg} for $4f$ production.
\end{sloppypar}

In detail, we start from the Born cross section based on the full set
of $\ggffff$ diagrams, which is parametrized as described in the
previous section and include the Higgs resonance with SM couplings.
Such lowest-order predictions, which we denote ``Born+Higgs''
below, have already been presented in \citere{Bredenstein:2004ef}.
In addition, we now dress the resulting cross section
with the off-shell Coulomb singularity,
\beq
\int\rd\si_{\mathrm{IBA}}^{\ggffff}=
\frac{1}{2s}\int\rd\Phi_{4f}\,
(1+\de_{\mathrm{coul}})|\M_{\mathrm{Born+Higgs}}^{\ggffff}|^2.
\eeq
The correction factor $\de_{\mathrm{coul}}$ for the Coulomb singularity
was calculated in \citere{Bardin:1993mc} to
\beq
\de_{\mathrm{coul}}=
\frac{\al(0)}{\bar\beta}\Im\left\{\ln
\left(\frac{\beta+\Delta-\bar\beta}{\beta+\Delta+\bar\beta}\right)\right\},
\eeq 
with the abbreviations
\beqar
\bar\beta &=& \frac{1}{s}\sqrt{s^2+(k_+^2)^2+(k_-^2)^2
-2sk_+^2-2sk_-^2-2k_+^2k_-^2},\nn\\
\beta &=& \sqrt{1-\frac{4(\MW^2-\ri\MW\GW)}{s}}, \qquad
\Delta = \frac{|k_+^2-k_-^2|}{s}.
\eeqar

\section{Treatment of soft and collinear photon emission}
\label{se:softcoll}

We calculate the real photonic corrections from the full lowest-order 
matrix element of the process \refeq{eq:bremsprocess}
without any expansion about the W-boson resonances.
They are calculated from the integral
\beq
\int\rd\si^{\ggffffg}=
\frac{1}{2s}\int\rd\Phi_{4f\ga}\, |\M^{\ggffffg}|^2 \, \Theta(\Phi_{4f\ga}),
\label{eq:realrc}
\eeq
where we have made the implementation of phase-space cuts explicit 
by including the step function $\Theta(\Phi_{4f\ga})$, which
is equal to 1 if an event passes the cuts and 0 otherwise.
Since we evaluate the real matrix element $\M^{\ggffffg}$ with
massless particles, the phase-space integral diverges in the soft
and collinear regions, where the emitted photon is either soft or
collinear to an outgoing external charged fermion.
In these regions we reintroduce a formally
infinitesimal photon mass $\la$ and small fermion masses $m_f$ as
regulators.

To this end, we apply two different methods:
the dipole subtraction and the (two-cutoff)
phase-space slicing methods. 
In the case of collinear-safe observables we closely follow 
the approach of \citere{Denner:2000bj} and only give 
a brief description in \refse{se:coll-safe} since the procedure 
is very similar to the $\Pep\Pem$ case. 
In \refse{se:non-coll-safe} we describe how the two methods are extended to 
non-collinear-safe observables. 

\subsection{Collinear-safe observables}
\label{se:coll-safe}

\subsubsection{Phase-space slicing}

In the phase-space slicing approach the phase space is divided into 
regions where the integrand is finite and can, thus, be integrated 
numerically, and regions where the integrand becomes singular. 
In the singular regions the integration over the photon phase space 
is carried out analytically 
in the approximation that the photon is soft and/or collinear to a 
charged fermion.

The singular regions consist of two parts one of which contains 
a soft photon ($k^0<\Delta E$) and the other a photon 
that is collinear but not soft ($k^0>\Delta E$ and 
$\theta_{\ga f}<\Delta\theta$, where $\theta_{\ga f}$ is 
the angle between the photon and a charged fermion). 
Thus, the real corrections are decomposed according to
\beq
\int\rd\si^{\ggffffg}=
%\frac{1}{2s}\rd\Phi_{4f\ga}|\M^{\ggffffg}|^2 \, \Theta(\Phi_{4f\ga})=
\int\rd\si_{\soft} + \int\rd\si_{\coll} 
+ \int\rd\si_{\finite}^{\ggffffg},
\eeq
where the cutoff parameters $\De E$ and $\De\theta$ are defined
in the CM system of the incoming photons.
Both in the soft and collinear regions the squared matrix element 
$|\M^{\ggffffg}|^2$ factorizes into the 
squared lowest-order matrix element 
$|\M_{\Born}^{\ggffff}|^2$ and a universal factor containing the singularity. 
The five-particle phase space also factorizes into a 
four-particle phase space and a photon part, so that 
$\rd\si_{\soft}$ and $\rd\si_{\coll}$ can be integrated 
over the photon momentum. Taking over the results from 
\citere{Denner:2000bj} yields 
\beqar
\rd\si_{\soft} &=& 
\rd\si_{\Born}^{\ggffff} \, \Theta(\Phi_{4f}) \,
\frac{\al}{\pi}\sum^4_{i=1}\sum^4_{j=i+1}(-1)^{i+j}Q_iQ_j 
\left\{ 2\ln\left(\frac{2\Delta E}{\la}\right)
\left[1-\ln\left(\frac{s_{ij}}{m_im_j}\right)\right]
\right. 
\nn\\
&& \left. \quad {}
-\ln\left(\frac{4p^0_i p^0_j}{m_im_j}\right) 
+ \ln^2\left(\frac{2p^0_i}{m_i}\right) 
+ \ln^2\left(\frac{2p^0_j}{m_j}\right) + \frac{\pi^2}{3} 
+ \Li\left(1-\frac{4p^0_ip^0_j}{s_{ij}}\right)\right\}
\eeqar
and
\beq
\label{eq:coll}
\rd\si_{\coll} = 
\rd\si_{\Born}^{\ggffff} \, \Theta(\Phi_{4f}) \,
\frac{\al}{2\pi} \sum^4_{i=1}Q_i^2 
%\nn\\
%&& \qquad\qquad\times
\Biggl\{\Biggl[\frac{3}{2}
+2\ln\left(\frac{\Delta E}{p^0_i}\right)\Biggr]
\Biggl[1-2\ln\left(\frac{\Delta\theta\, p^0_i}{m_i}\right)\Biggr] 
+3-\frac{2\pi^2}{3}\Biggr\},
\eeq
where $Q_i$ and $m_i$ denote the relative electric charge and mass
of fermion $f_i$, respectively.
The step function $\Theta(\Phi_{4f})$ indicates that both 
$\rd\si_{\soft}$ and $\rd\si_{\coll}$ are defined on the
four-particle phase space of the lowest-order cross section,
so that the singular part 
\beq
\rd\si_{\sing}^{\ggffffg} = 
\rd\si_{\soft} + \rd\si_{\coll}
\eeq
can be locally combined with the singular part of the virtual corrections,
which are defined on the same phase space.
In the result $\rd\si_{\mathrm{virt+real,sing}}^{\ggffff}$ 
all dependences on the photon and fermion masses 
%($\ln\la,\ln m_i$) 
cancel. 

\begin{figure}
\setlength{\unitlength}{1cm}
\centerline{
\begin{picture}(7.7,8)
\put(-1.7,-14.5){\includegraphics{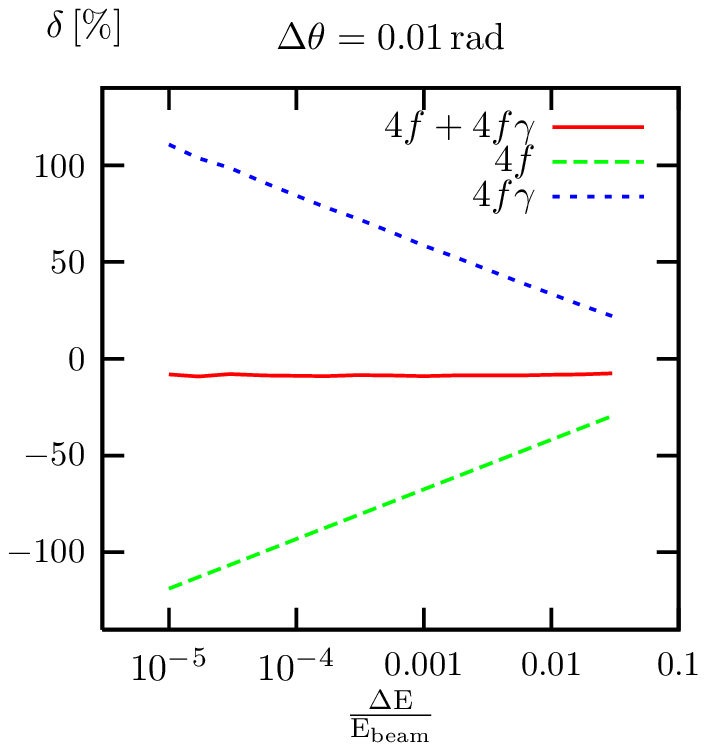}}
\end{picture}
\begin{picture}(7.5,8)
\put(-1.7,-14.5){\includegraphics{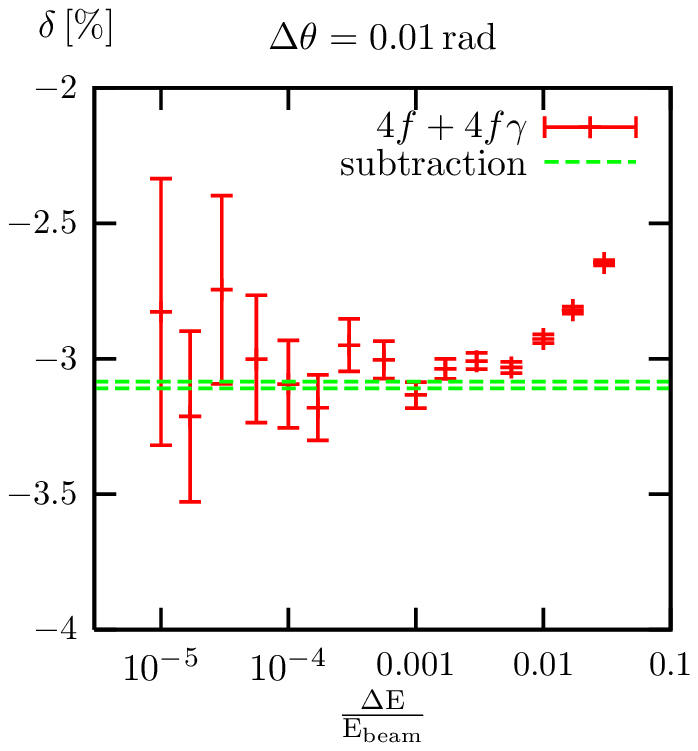}}
\end{picture}}
\caption{Dependence of the corrections on the energy cutoff 
in the slicing approach for the process $\ga\ga\to\Pne\Pep\Pd\Pubar$ 
at $\sqrt{s_{\ga\ga}}=500 \GeV$. For comparison the corresponding 
result obtained with the dipole subtraction method is shown as a $1\si$ band 
in the plot on the r.h.s.}
\label{fig:slienergy}
%\end{figure}
\vspace*{1.5em}
%\begin{figure}
\setlength{\unitlength}{1cm}
\centerline{
\begin{picture}(7.7,8)
\put(-1.7,-14.5){\includegraphics{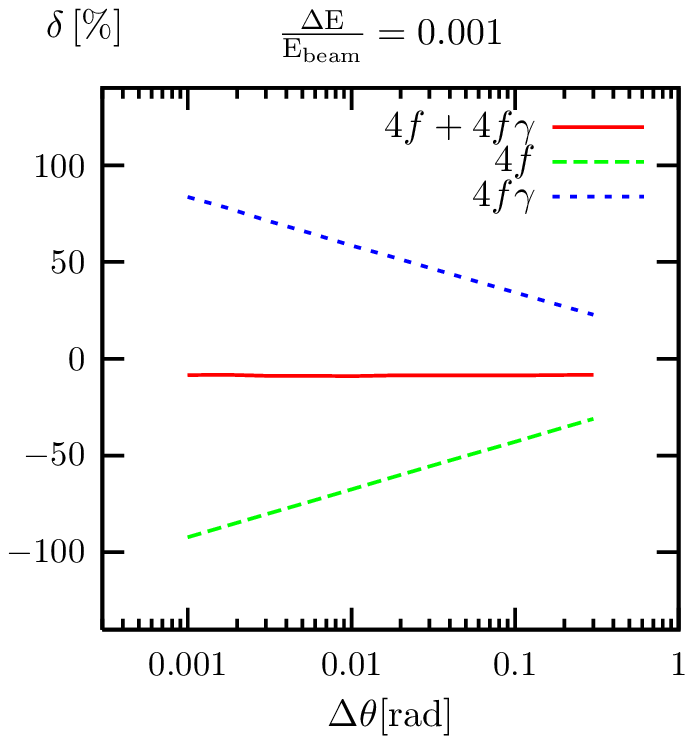}}
\end{picture}
\begin{picture}(7.5,8)
\put(-1.7,-14.5){\includegraphics{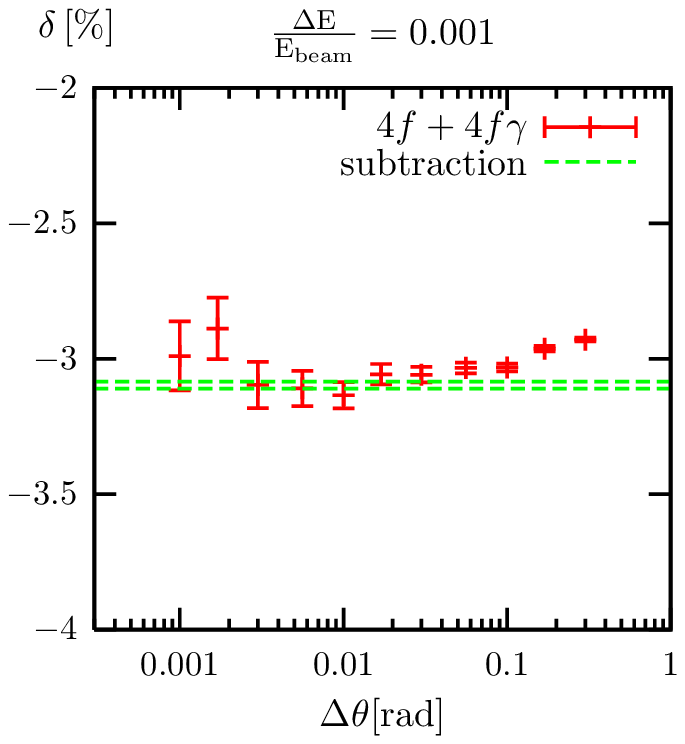}}
\end{picture}}
\caption{Dependence of the corrections on the angular cutoff 
in the slicing approach for the process $\ga\ga\to\Pne\Pep\Pd\Pubar$ 
at $\sqrt{s_{\ga\ga}}=500 \GeV$. For comparison the corresponding 
result obtained with the subtraction method is shown as a $1\si$ band 
in the plot on the r.h.s.}
\label{fig:sliangle}
\end{figure}

While $\rd\si_{\mathrm{virt+real,sing}}^{\ggffff}$ depends on the 
cutoff parameters $\Delta E$ and $\Delta\theta$ 
analytically, the finite real corrections 
$\int\rd\si_{\finite}^{\ggffffg}$
only show this dependence upon the cuts in the numerical integration.
Nevertheless, the cutoff
dependence has to cancel in the full result in the limit 
$\Delta E, \Delta\theta\to 0$.
This is illustrated on the l.h.s.\ of \reffis{fig:slienergy} and 
\ref{fig:sliangle} where the relative correction factor 
$\de=\si/\si_{\Born}-1$
of the $4f$ part 
($\int\rd\si_{\mathrm{virt,finite,DPA}}^{\ggffff} 
+\int\rd\si_{\mathrm{virt+real,sing}}^{\ggffff}$) and of 
the $4f\ga$ part $\int\rd\si_{\finite}^{\ggffffg}$ 
is shown as a function of the cutoff parameters $\Delta E$ and $\Delta\theta$.
The cancellations of the cutoff dependence of the two contributions 
is shown on a smaller scale on the r.h.s.\ of \reffis{fig:slienergy} and 
\ref{fig:sliangle}. 
While terms of $\Ord(\Delta E/E_{\mathrm{beam}})$ and $\Ord(\Delta\theta)$ 
become visible for large values of the cutoff parameters, for smaller values 
a plateau is reached. The integration error increases with decreasing
cutoff values, until for too small values the integration 
error is usually underestimated.
As a result, we decided to take 
$\Delta E/E_{\mathrm{beam}}=10^{-3}$ and $\Delta\theta=10^{-2}$
as default values.

\subsubsection{Dipole subtraction method}

In a subtraction method an auxiliary function is constructed 
that contains the same singularities as the real corrections. 
Subtracting this function from the real corrections, this difference can be 
integrated numerically. The next step is 
to perform the singular integration of
the auxiliary function over the photon momentum analytically
and to readd the result to the virtual corrections.
In our case where 
soft and collinear singularities originate from final-state
radiation only, 
the soft and collinear singularities completely cancel
against their counterparts in the virtual corrections
for collinear-safe observables.

In the dipole subtraction method \cite{Dittmaier:1999mb,Roth:1999kk},
which was originally proposed for QCD \cite{Catani:1996jh},
the auxiliary function consists of different contributions 
labelled by all ordered combinations of two charged fermions $i$ and $j$, 
which are called emitter and spectator. These contributions contain the 
singularities connected with the emitter $i$. 
Since there are only charged particles in the final state in $\ggffff$, 
the situation is simpler than for $\eeffff$. 
%because a photon emitted from an initial-state fermion changes 
%the kinematics of the subsequent hard scattering
%process and not just the momentum of two fermions. 
Explicitly the auxiliary function, which is subtracted from the
spin-summed squared bremsstrahlung matrix element, reads
\beqar
|\M_{\sub}|^2 &=& \sum_{i,j=1\atop i\ne j}^4|\M_{\sub,ij}|^2, 
\nn\\
|\M_{\sub,ij}(\Phi_{4f\ga})|^2 &=& -(-1)^{i+j}Q_iQ_je^2
g^{(\sub)}_{ij}(p_i,p_j,k)|\M_{\Born}
^{\ggffff}(\tilde\Phi_{4f,ij})|^2.
\label{eq:msub}
\eeqar
Adopting the formulation of \citere{Dittmaier:1999mb}%
\footnote{The formulation of \citere{Roth:1999kk} differs from that
by the regular factor $1/(1-y_{ij})$ in Eq.~\refeq{eq:gsubij}, so that 
the readded singular contributions of \citeres{Dittmaier:1999mb} and
\cite{Roth:1999kk} differ by non-singular finite parts.},
the soft and collinear divergences are contained in the function
\beq
g^{(\sub)}_{ij}(p_i,p_j,k)
=\frac{1}{(p_ik)(1-y_{ij})}\left[\frac{2}{1-z_{ij}(1-y_{ij})}-1-z_{ij}\right]
\label{eq:gsubij}
\eeq
with
\beq
y_{ij}=\frac{p_ik}{p_ip_j+p_ik+p_jk},
\qquad z_{ij}=\frac{p_ip_j}{p_ip_j+p_jk}.
\label{eq:yijzij}
\eeq
The embedding of the $4f$ phase space $\tilde\Phi_{4f,ij}$ 
into the $4f\ga$ phase space $\Phi_{4f\ga}$ 
is defined as 
\beq
\tilde p_i^{\mu}=p_i^{\mu}+k^{\mu}-\frac{y_{ij}}{1-y_{ij}}p_j^{\mu},
\qquad \tilde p_j^{\mu}=\frac{1}{1-y_{ij}}p_j^{\mu},
\label{eq:ptilde}
\eeq
with all other momenta unchanged, $\tilde p_k=p_k, k\ne i,j$.
Subtracting the auxiliary function 
from the real corrections enables us to carry out the numerical integration,
\beq
\int\rd\si_{\finite}^{\ggffffg}=\frac{1}{2s}\int\rd\Phi_{4f\ga}
\left[|\M^{\ggffffg}|^2\Theta(\Phi_{4f\ga})
-\sum_{i,j=1 \atop i\ne j}^4|\M_{\sub,ij}|^2\Theta(\tilde\Phi_{4f,ij})\right],
\label{eq:mdiff}
\eeq
which does not contain any soft or collinear divergences by 
construction of $|\M_{\sub}|^2$ for collinear-safe observables.
In this context,
it is important to notice the different arguments of the step
functions $\Theta$ which account for phase-space cuts. 
Since for a generic point in $4f\gamma$ phase space
each $ij$ contribution corresponds to a different point in phase space, 
there is in general no correlation between the values of the different
step functions.
For collinear-safe observables, however, we have 
$\Theta(\Phi_{4f\ga})=\Theta(\tilde\Phi_{4f,ij})$ in the soft region
$(k\to0)$ and in the region where the photon momentum $k$ is nearly
collinear to the emitter momentum $p_i$ ($p_i k\to 0$).
The collinear safety can, e.g., be enforced by photon recombination, 
as discussed in the next section in more detail.

In order to combine the 
subtraction function with the virtual correction,
it has to be integrated over the photon momentum, yielding 
\beq
\label{eq:subsing}
\int\rd\si_{\sing}^{\ggffffg}=-\frac{\al}{2\pi}
\sum^4_{{i,j=1}\atop{i\neq j}}(-1)^{i+j}Q_iQ_j\frac{1}{2s}
\int\rd\Phi_{4f}\, G^{(\sub)}_{ij}(s_{ij})|\M_{\Born}^{\ggffff}(\Phi_{4f})|^2
\Theta(\Phi_{4f}).
\eeq
The singularities are contained in the function 
\beq
G^{(\sub)}_{ij}(s_{ij})=
\L(s_{ij},m_i^2)-\frac{\pi^2}{3}+\frac{3}{2}
\label{eq:Gsub}
\eeq
with 
\beq
\L(s_{ij},m_i^2)
=\ln\left(\frac{m_i^2}{s_{ij}}\right)\ln\left(\frac{\la^2}
{s_{ij}}\right)
+\ln\left(\frac{\la^2}{s_{ij}}\right)-\frac{1}{2}\ln^2\left(\frac{m_i^2}
{s_{ij}}\right)+\frac{1}{2}\ln\left(\frac{m_i^2}{s_{ij}}\right).
\label{eq:L}
\eeq
We have numerically checked that these soft and collinear divergences 
are completely cancelled by their counterparts in the
virtual correction.

\subsection{Non-collinear-safe observables}
\label{se:non-coll-safe}

In the previous sections the matching of real and virtual corrections 
was described for collinear-safe observables. 
We speak of collinear-safe observables
if a nearly collinear system of a charged fermion and a photon is treated 
inclusively, i.e.\ if phase-space selection cuts 
(or histogram bins of distributions)
depend only on the sum $p_i+k$ of the nearly collinear
fermion and photon momenta. In this case the energy fraction 
\beq
z_i=\frac{p^0_i}{p^0_i+k^0}
\label{eq:zi}
\eeq
of a charged fermion $f_i$ after emitting a photon
in a sufficiently small cone around its direction of flight
is fully integrated over, because it is not constrained by any 
phase-space cut 
(or histogram bin selection in distributions). 
Thus, the KLN theorem \cite{Kinoshita:1962ur} 
guarantees that all singularities connected with final-state radiation cancel 
between the virtual and real corrections, even though they are defined 
on different phase spaces.
A sufficient inclusiveness is, e.g., achieved by the photon recombination
described in \refse{se:input}, which treats outgoing charged fermions
and photons as one quasi-particle if they are very close in angle.

In the previous section we could, therefore, integrate the subtraction 
function $|\M_{\sub}|^2$ and the slicing contribution $\rd\si_{\coll}$ 
over $z_i$ analytically. 
In this section we are concerned with non-collinear-safe observables, i.e.\ 
the fermion--photon system is not treated inclusively and fermion-mass 
singularities can become visible. As the integration over $z_i$ 
now is constrained by phase-space cuts (or histogram bins), we have 
to modify the methods described in the previous section in such a way
that the integration over $z_i$ is part of the numerical phase-space
integration.

\subsubsection{Phase-space slicing}

In the slicing method the procedure is straightforward. The numerical 
integration over $z=z_i$ in the collinear parts reads 
\beqar
\label{eq:coll_z}
\rd\si_{\coll}&=&
\rd\si_{\Born}^{\ggffff}(\tilde\Phi_{4f}) \,
\frac{\al}{2\pi}\sum^4_{i=1}Q_i^2
\int_0^{1-\Delta E/\tilde p^0_i}\rd z\,
\Theta\Big(p_i=z\tilde p_i, k=(1-z)\tilde p_i,\{\tilde p_{j\ne i}\}\Big)
\nn\\
&& \hspace*{4em}{} \times
\Biggl\{P_{ff}(z)\left[
2\ln\left(\frac{\Delta\theta\,\tilde p^0_i}{m_i}z\right)
-1\right]+(1-z)\Biggr\},
\eeqar
with the splitting function
\beq
P_{ff}(z)=\frac{1+z^2}{1-z}.
\label{eq:Pff}
\eeq
The Born cross section and the logarithm still depend on the momenta of 
the $4f$ phase space $\tilde\Phi_{4f}$ which are labelled $\tilde p_i$. 
In the cut and recombination function $\Theta$, however, 
the momentum $\tilde p_i$ of the fermion $i$ (before photon emission)
is distributed to the fermion momentum $p_i$ and the photon momentum $k$. 
For collinear-safe observables, as e.g.\ achieved by photon recombination,
the $\Theta$ function effectively only depends on the sum 
$p_i+k=\tilde p_i$ of the collinear momenta, which is independent of $z$.
In this case, the $\Theta$ function becomes $\Theta(\tilde\Phi_{4f})$,
and the $z$-integration can be easily carried out analytically
yielding Eq.~\refeq{eq:coll}.

\subsubsection{Dipole subtraction method}
\label{se:dps-ncs}

In the case of the dipole subtraction method the generalization to 
non-collinear-safe observables is more complicated than in the 
slicing approach, since the integration over the photon momentum 
is more involved. 
Here, we collect the formulas relevant for our calculation. Details on
their derivation are given in \refapp{app:sub}.

In order to keep the information on the energy fraction $z$
in each part of the subtraction function, the finite part of the
real corrections is modified to
\beqar
\int\rd\si_{\finite}^{\ggffffg} &=& \frac{1}{2s}\int\rd\Phi_{4f\ga}
\Biggl[|\M^{\ggffffg}|^2\Theta(\Phi_{4f\ga})
\nn\\
&& {}
-\sum_{i,j=1 \atop i\ne j}^4|\M_{\sub,ij}|^2
\Theta\Big(p_i=z_{ij}\tilde p_i,k=(1-z_{ij})\tilde p_i,
\{\tilde p_{k\ne i}\}\Big)\Biggr].
\eeqar
It is easily seen that the variable $z_{ij}$, which is defined in 
Eq.~\refeq{eq:yijzij}, plays the role
of the energy fraction $z_i$ in the collinear limit for each
dipole $ij$.
Again, in the collinear-safe case the $\Theta$ functions of the
subtraction function depend only on the sums $p_i+k=\tilde p_i$
of collinear momenta; in this case we recover Eq.~\refeq{eq:mdiff}.

In the integration of the subtraction function over the photon
phase space, we now have to leave the integrations over $z_{ij}$ open.
The resulting $z_{ij}$ dependence of the integrand
is most conveniently described
with a $[...]_+$ prescription%
\footnote{We use the definition
$\int_0^1\rd x\, \left[f(x)\right]_+ g(x) \equiv
\int_0^1\rd x\, f(x) \left[g(x)-g(1)\right]$.},
which separates the soft singularity at
$z_{ij}=1$. The endpoint part at $z_{ij}=1$, which results from
the full integration over $z_{ij}$, exactly corresponds to the
contribution of $G^{(\sub)}_{ij}(\tilde s_{ij})$ for the
collinear-safe case, as given in Eq.~\refeq{eq:Gsub},
where $\tilde s_{ij}=2\tilde p_i\tilde p_j$.
The continuum part in $z_{ij}$ involves an integral over
$\left[\bar{\cal G}^{(\sub)}_{ij}(\tilde s_{ij},z_{ij})\right]_+$ with 
\beq
\bar{\cal G}^{(\sub)}_{ij}(\tilde s_{ij},z)
=P_{ff}(z)\left[\ln\left(\frac{\tilde s_{ij}z}{m_i^2}\right)-1\right]
+(1+z)\ln(1-z)+(1-z).
\label{eq:bcGsub}
\eeq
The total integrated subtraction part explicitly reads
\beqar
\int\rd\si_{\sing}^{\ggffffg} &=& -\frac{\al}{2\pi}
\sum^4_{{i,j=1}\atop{i\neq j}}(-1)^{i+j}Q_iQ_j\frac{1}{2s}
\int\rd\tilde\Phi_{4f,ij}\,|\M_{\Born}^{\ggffff}(\tilde\Phi_{4f,ij})|^2 
\Biggl\{G^{(\sub)}_{ij}(\tilde s_{ij}) \Theta(\tilde\Phi_{4f,ij})
\nn\\
&& {} 
+\int_0^1\rd z \,
\left[\bar{\cal G}^{(\sub)}_{ij}(\tilde s_{ij},z)\right]_+
\Theta\Big(p_i=z\tilde p_i, k=(1-z)\tilde p_i,\{\tilde p_{k\ne i}\}\Big)
\Biggr\}.
\eeqar
Owing to the $[...]_+$ prescription, the continuum part is zero if
the full integration over $z$ is carried out, thereby recovering the
collinear-safe case \refeq{eq:subsing}.

\section{Numerical results}
\label{se:numerics}

\subsection{Input parameters and setup}
\label{se:input}

We use the following set of input parameters \cite{Hagiwara:fs}:
\beqar
\begin{array}[b]{r@{\,}lr@{\,}lr@{\,}l}
G_{\mu} &= 1.16639\times 10^{-5}\GeV^{-2}, \qquad &
\al(0) &= 1/137.03599976, \qquad & \alpha_{\mathrm{s}} &= 0.1172,\\
\MW &= 80.423\GeV, & \GW &= 2.118\GeV, && \\
\MZ &= 91.1876\GeV, & \GZ &= 2.4952\GeV, && \\
%M_{\PH} &= 170\GeV, & \GH &= 0.3834\GeV, && \\
\Me &= 0.510998902\times 10^{-3}\GeV, & \Mmy &= 0.105658357 \GeV, 
&\Mta &= 1.77699\GeV,\\
\Mu &= 0.066 \GeV, & \Mc &=1.2 \GeV, & \Mt &= 174.3\GeV, \\
\Md &= 0.066 \GeV, & \Ms &=0.15 \GeV, & \Mb &= 4.3\GeV.
\end{array}
\eeqar
If not stated otherwise, the Higgs mass is $\MH=170\GeV$. 
In some cases we alternatively use $\MH=130\GeV$.
The corresponding values for the Higgs-boson decay width $\GH$,
which have been obtained with the program
{\HDECAY} \cite{Djouadi:1997yw}, are given by
\beq
\GH\left(\MH=170\GeV\right) = 0.3834\GeV, \qquad
\GH\left(\MH=130\GeV\right) = 0.004995\GeV.
\eeq
We set the quark-mixing matrix to the unit matrix throughout,
but in the limit of massless external fermions a
non-trivial quark-mixing matrix can be included by a simple
rescaling of the cross sections.

Furthermore, we apply a set of recombination and separation cuts:
\begin{enumerate}
\item
Bremsstrahlung
photons that are closer than $5^{\circ}$ to a charged fermion 
or have less energy than $1\GeV$ 
are recombined with the charged fermion that is closest in angle. 
This means that in this case
before evaluating distributions or applying 
phase-space cuts the momenta of the photon and the fermion are added and 
associated with the fermion, while the photon is discarded.
\item
The following separation cuts are applied to the momenta defined 
after a possible recombination:
\beqar
\label{eq:cuts}
\begin{array}[b]{r@{\,}lr@{\,}lr@{\,}lr@{\,}l}
E_l&>10\GeV, \qquad & \theta(l,\mathrm{beam})&>5^{\circ}, & \qquad
\theta(l,l')&>5^{\circ}, & \theta(l,q)&>5^{\circ}, \\
E_q&>10\GeV, & \theta(q,\mathrm{beam})&>5^{\circ}, & 
m(q,q')&>10\GeV,
\end{array}
\eeqar
where an obvious notation for energies $E_{\dots}$, angles $\theta(\dots)$, 
and invariant masses $m(\dots)$ for leptons $l$ and quarks $q$ is used.
\end{enumerate}

The separation cuts and input parameters are the same as  
in \citere{Bredenstein:2004ef}%
\footnote{There is a misprint in Eq.~(6.1) of
\citere{Bredenstein:2004ef}. The value for $\alpha_s$ is supposed to 
be $\alpha_s=0.1172$ and not $\alpha_s=1.1172$.} 
for the processes $\ggffff$,
so that we reproduce the Born cross sections that we calculated there.
In particular, we exclude forward and backward scattered charged fermions,
because they cause collinear singularities.
While for final-state quarks these singularities
signal a non-perturbative regime,
for leptons they are in principle
cured by finite-mass effects. However,
we exclude this region by demanding
that leptons appear in the detector with finite production
angle and energy.
Compared to \citere{Denner:2000bj} we use different 
recombination cuts, because, in contrast to $\Pep\Pem$ collisions, 
%an invariant mass cannot be used to trigger the recombination. 
the recombination criterion based on invariant masses does not lead to
collinear-safe observables.
This is due to the collinear singularity that arises if a charged 
fermion is collinear to the beam. Even though an appropriate cut 
on the angle between charged fermions and the beam is imposed, 
it might happen that a photon with relatively high energy is 
recombined with a low-energy fermion that is close to the beam. 
Thus, after recombination, the fermion almost follows the direction 
of the photon and is not affected by the angular cut. 
Such events are avoided by taking a recombination condition based
on the angle.

For the evaluation of the lowest-order matrix elements of $\ggffff$
and $\ggffffg$, we
use the fixed-width scheme, in which the gauge-boson width is 
introduced in all (i.e.\ time- and space-like) propagators.
As argued in Section~2.4 of \citere{Bredenstein:2004ef}, this scheme
does not break gauge invariance for reactions $\ggffff(+\ga)$ 
with massless external fermions.          

The photon spectrum is accounted for by using the parametrization 
of the program CompAZ \cite{Zarnecki:2002qr}, as described in 
Section~5 of \citere{Bredenstein:2004ef}.
In order to distinguish the cases with and without convolution over
the photon spectrum, we write $\sqrt{s_{\Pe\Pe}}$ and $\sqrt{s_{\ga\ga}}$
for the CM energies in these cases, respectively.

In the numerical integration we generate $2\cdot 10^7$ events 
for the plots showing the integrated cross sections, and $5\cdot 10^7$ events 
for distributions and for the integrated cross sections in 
\refta{ta:intcross}.
If not stated otherwise,
the shown results are based on the subtraction method, but have been
cross-checked with the slicing approach. Moreover, we have additionally
checked 
most of the results by reproducing them within statistical uncertainties
with our second independent Monte Carlo generator.

\subsection{Integrated cross sections}

\begin{table}
\centerline{
\begin{tabular}{|c|c|c|c|c|c|}
\hline
& & \multicolumn{2}{c|}{$\sigma[\fb]$} & $\sigma_{\Born}[\fb]$ & \\
\hline
CM energy & final state & subtraction & slicing & & (sub--sli)/sli\\
\hline
\hline
 & $                              \Pne\Pep\Pmum\Pnmubar         $
 &  581.403(67)
 &  581.41(16)
 &  575.628(64)
 &    0.00(3)
   \%
 \\\cline{2-6}
$\sqrt{s_{\ga\ga}}=200\GeV$
 & $                              \Pne\Pep\Pd\Pubar             $
 & 1734.02(23)
 & 1735.26(43)
 & 1716.10(22)
 &   --\,0.07(3)
   \%
 \\\cline{2-6}
without $\ga$ spectrum
 & $                              \Pu\Pdbar\Pem\Pnebar          $
 & 1734.24(23)
 & 1734.32(43)
 & 1716.06(22)
 &    0.00(3)
   \%
 \\\cline{2-6}
 & $                              \Pu\Pdbar\Ps\Pcbar            $
 & 4931.01(76)
 & 4935.0(1.0)
 & 4878.67(73)
 &   --\,0.08(3)
   \%
\\\hline \hline
 & $                              \Pne\Pep\Pmum\Pnmubar         $
 &  801.21(11)
 &  801.57(20)
 &  826.620(91)
 &   --\,0.05(3)
   \%
 \\\cline{2-6}
$\sqrt{s_{\ga\ga}}=500\GeV$
 & $                              \Pne\Pep\Pd\Pubar             $
 & 2278.50(34)
 & 2279.96(51)
 & 2351.37(30)
 &   --\,0.06(3)
   \%
 \\\cline{2-6}
without $\ga$ spectrum
 & $                              \Pu\Pdbar\Pem\Pnebar          $
 & 2278.45(34)
 & 2278.84(48)
 & 2351.39(30)
 &   --\,0.02(3)
   \%
 \\\cline{2-6}
 & $                              \Pu\Pdbar\Ps\Pcbar            $
 & 6452.2(1.0)
 & 6452.8(1.2)
 & 6662.25(96)
 &   --\,0.01(2)
   \%
\\\hline \hline
 & $                              \Pne\Pep\Pmum\Pnmubar         $
 &  696.25(15)
 &  696.68(17)
 &  746.995(93)
 &   --\,0.06(3)
   \%
 \\\cline{2-6}
$\sqrt{s_{\ga\ga}}=1000\GeV$
 & $                              \Pne\Pep\Pd\Pubar             $
 & 1836.31(43)
 & 1836.96(45)
 & 1979.92(29)
 &   --\,0.04(3)
   \%
 \\\cline{2-6}
without $\ga$ spectrum
 & $                              \Pu\Pdbar\Pem\Pnebar          $
 & 1836.37(42)
 & 1836.95(42)
 & 1979.95(29)
 &   --\,0.03(3)
   \%
 \\\cline{2-6}
 & $                              \Pu\Pdbar\Ps\Pcbar            $
 & 4892.2(1.2)
 & 4891.4(1.1)
 & 5300.97(90)
 &    0.02(3)
   \%
\\\hline \hline
 & $                              \Pne\Pep\Pmum\Pnmubar         $
 &    0.073205(44)
 &    0.073205(44)
 &    0.072009(44)
 &    0%.00(9) \%
 \\\cline{2-6}
$\sqrt{s_{\Pe\Pe}}=200\GeV$
 & $                              \Pne\Pep\Pd\Pubar             $
 &    0.33129(21)
 &    0.33129(21)
 &    0.32601(21)
 &    0%.00(9) \%
 \\\cline{2-6}
with $\ga$ spectrum
 & $                              \Pu\Pdbar\Pem\Pnebar          $
 &    0.39204(25)
 &    0.39204(25)
 &    0.38593(24)
 &    0%.00(9) \%
 \\\cline{2-6}
 & $                              \Pu\Pdbar\Ps\Pcbar            $
 &    1.24460(79)
 &    1.24460(79)
 &    1.22537(78)
 &    0%.00(9) \%
\\\hline \hline
 & $                              \Pne\Pep\Pmum\Pnmubar         $
 &  190.757(60)
 &  190.835(96)
 &  190.816(45)
 &   --\,0.04(6)
   \%
 \\\cline{2-6}
$\sqrt{s_{\Pe\Pe}}=500\GeV$
 & $                              \Pne\Pep\Pd\Pubar             $
 &  559.18(18)
 &  559.63(24)
 &  558.50(14)
 &   --\,0.08(5)
   \%
 \\\cline{2-6}
with $\ga$ spectrum
 & $                              \Pu\Pdbar\Pem\Pnebar          $
 &  564.58(18)
 &  564.79(25)
 &  565.05(14)
 &   --\,0.04(5)
   \%
 \\\cline{2-6}
 & $                              \Pu\Pdbar\Ps\Pcbar            $
 & 1604.92(54)
 & 1605.60(59)
 & 1603.80(45)
 &   --\,0.04(5)
   \%
\\\hline \hline
 & $                              \Pne\Pep\Pmum\Pnmubar         $
 &  165.759(91)
 &  165.604(81)
 &  170.588(41)
 &    0.09(7)
   \%
 \\\cline{2-6}
$\sqrt{s_{\Pe\Pe}}=1000\GeV$
 & $                              \Pne\Pep\Pd\Pubar             $
 &  461.02(20)
 &  461.34(23)
 &  474.81(12)
 &   --\,0.07(7)
   \%
 \\\cline{2-6}
with $\ga$ spectrum
 & $                              \Pu\Pdbar\Pem\Pnebar          $
 &  472.10(19)
 &  471.61(24)
 &  485.65(13)
 &    0.10(7)
   \%
 \\\cline{2-6}
 & $                              \Pu\Pdbar\Ps\Pcbar            $
 & 1296.49(52)
 & 1295.29(62)
 & 1335.13(38)
 &    0.09(6)
   \%
\\\hline
\end{tabular}}
\caption{Integrated cross sections for different final states and energies 
with and without convolution over the photon spectrum. The third 
column shows the result obtained with the subtraction method and the 
fourth with the slicing method. The last two columns show the Born cross 
section and the relative difference between subtraction and slicing.}
\label{ta:intcross}
\end{table}
In \refta{ta:intcross} we present a survey of integrated cross sections 
for a leptonic, a hadronic, and two semi-leptonic final states, 
as obtained with the subtraction and slicing methods.
The cross sections of the semi-leptonic final states differ because 
of the effective polarizations of the photons resulting from the Compton 
backscattering 
(cf.\ Section~6.3 of \citere{Bredenstein:2004ef}). 
Final states that differ only in the fermion generation 
(i.e.\ in their mass values)
receive the same radiative corrections, since our predictions are
based on the massless limit for the external fermions and
mass singularities cancel after performing a photon recombination.
The results obtained with the two methods for treating the real corrections, 
subtraction (``sub'') and slicing (``sli''), are in good agreement. 
Note that they both are implemented in the same Monte Carlo generator, 
which, thus, yields identical results for $\sqrt{s_{\ga\ga}}<170\GeV$
where the IBA is used. This is the reason why the ``sub'' and ``sli''
numbers are identical in the case of 
$\sqrt{s_{\Pe\Pe}}=200\GeV$ with $\ga$ spectrum, where only the range
$\sqrt{s_{\ga\ga}}<170\GeV$ is relevant in the convolution.

\begin{figure}
\setlength{\unitlength}{1cm}
\centerline{
\begin{picture}(7,8)
\put(-1.7,-11.5){\includegraphics{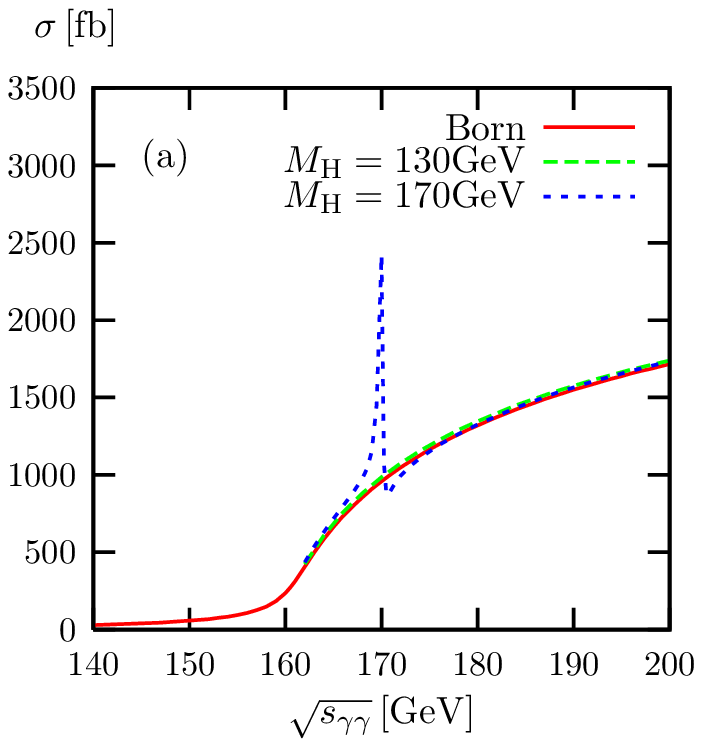}}
\end{picture}
\begin{picture}(7,8)
\put(-1.7,-11.5){\includegraphics{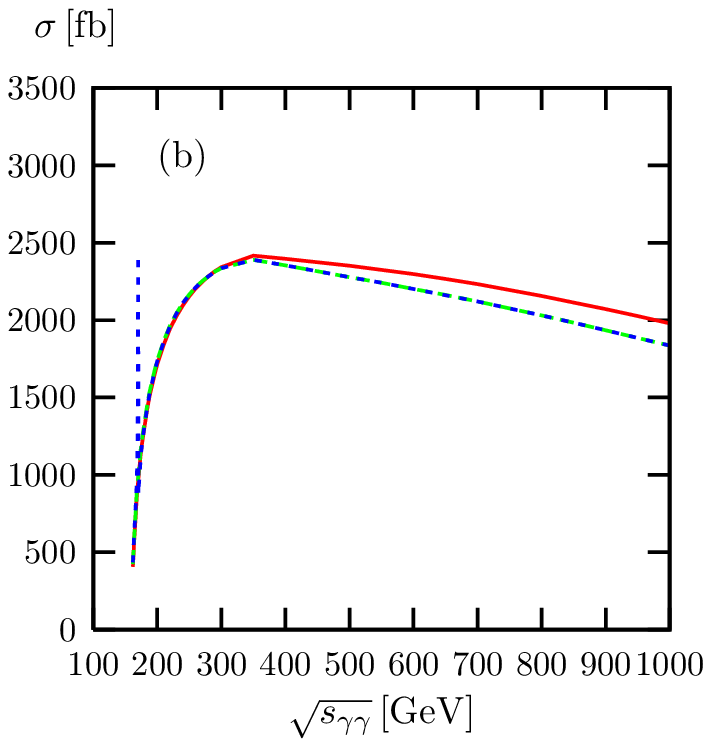}}
\end{picture} }
\centerline{
\begin{picture}(7,8)
\put(-1.7,-10.5){\includegraphics{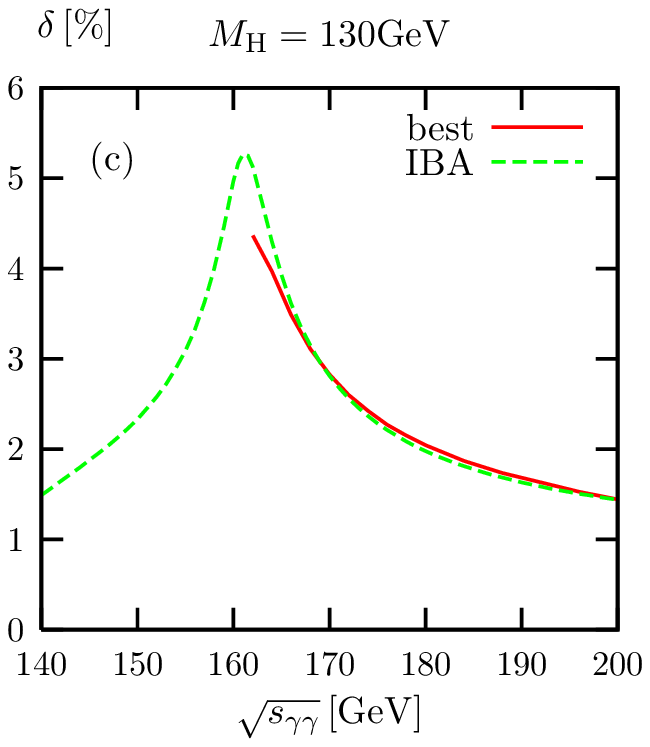}}
\end{picture}
\begin{picture}(7,8)
\put(-1.7,-10.5){\includegraphics{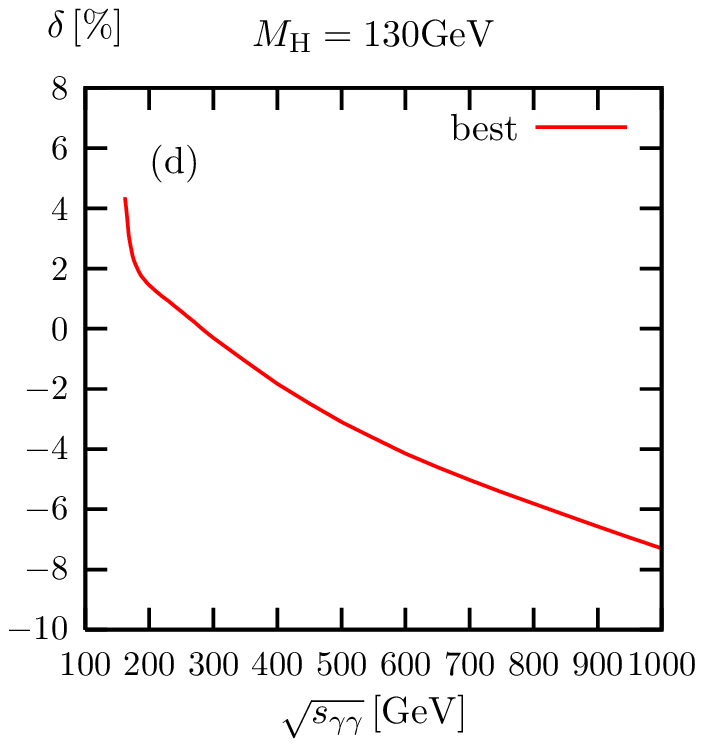}}
\end{picture} }
\centerline{
\begin{picture}(7,5)
\put(-1.7,-12.5){\includegraphics{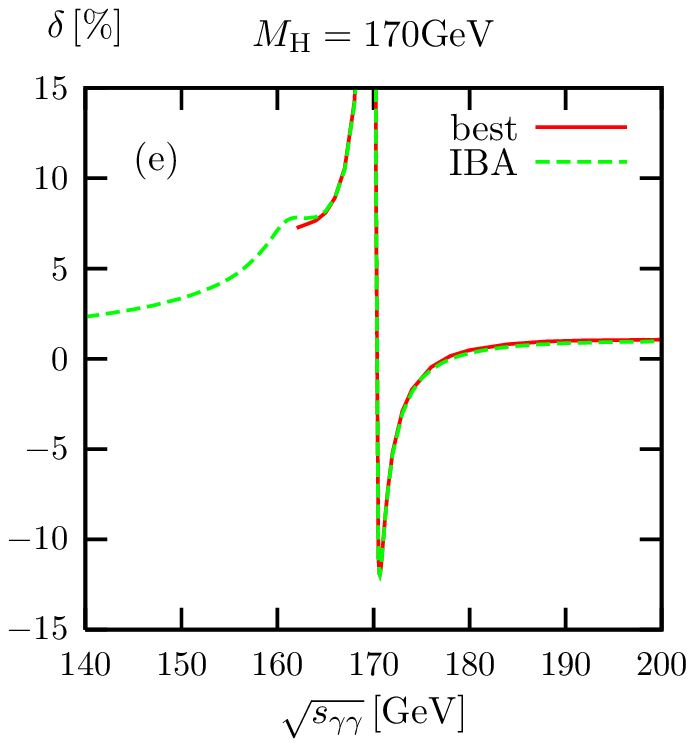}}
\put(.3,-3){\includegraphics{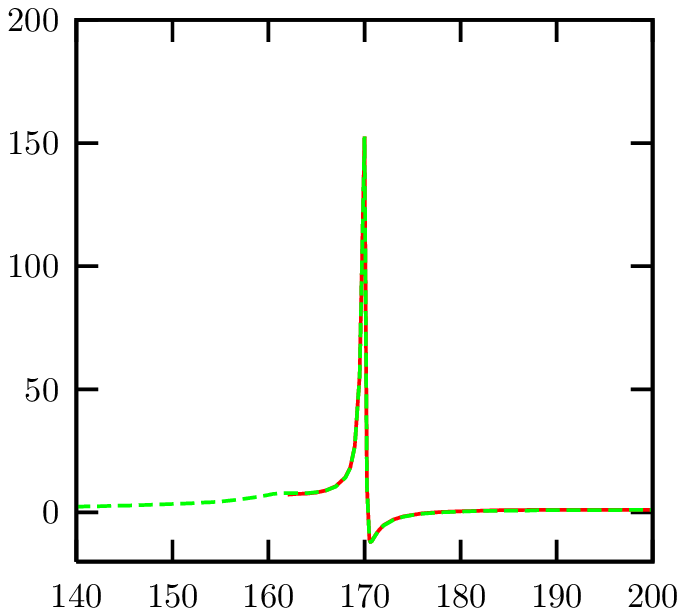}}
\end{picture}
\begin{picture}(7,5)
\put(-1.7,-12.5){\includegraphics{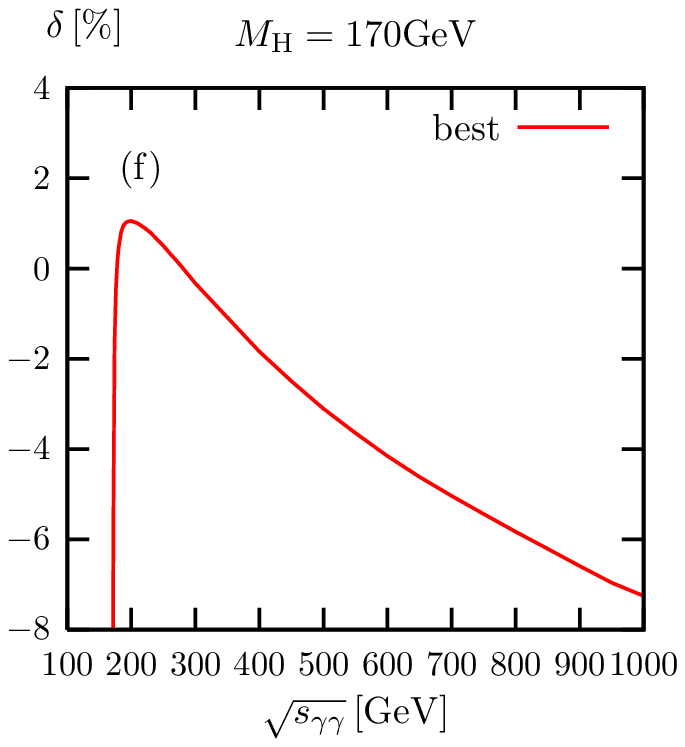}}
\end{picture} }
\caption{Integrated cross section for $\ga\ga\to\Pne\Pep\Pd\Pubar$
(the two upper plots) and relative radiative 
corrections (the four lower plots) 
without convolution over 
the photon spectrum for Higgs masses $\MH=130\GeV$ and $170\GeV$.}
\label{fig:sqrts.nospec}
\end{figure}

In \reffi{fig:sqrts.nospec} the integrated cross section 
for $\ga\ga\to\Pne\Pep\Pd\Pubar$ including 
radiative corrections is compared with the Born cross section as 
a function of the CM energy for monochromatic photon beams.
The ``best'' curves correspond to the ${\cal O}(\alpha)$-corrected
cross sections.
A Higgs boson of $\MH=170\GeV$ 
produces a sharp peak in the cross section at $\sqrt{s_{\ga\ga}}=170\GeV$,
while for larger energies the 
corrections are almost independent of the Higgs mass. 
The relative corrections 
$\de=\si/\si_{\Born}-1$ in the four lower plots of 
\reffi{fig:sqrts.nospec}
behave roughly like the corrections to on-shell
W-pair production \cite{Denner:1994ca,Denner:1995jv,Jikia:1996uu}. 
Close to the W-pair production threshold the corrections are 
dominated by the Coulomb singularity. 
For higher energies the corrections decrease until they reach 
about $-7 \%$ at $1 \TeV$. In this region they are dominated by 
large logarithms from the Regge and Sudakov domains. 

In \reffi{fig:sqrts.nospec}(c) we also show 
the comparison with the IBA for a Higgs mass of $\MH=130\GeV$.
Since close to the W-pair production threshold 
the bulk of the corrections is due to the Coulomb singularity and
since there are no other pronounced corrections, 
the agreement between the two curves is quite good. 
The very good agreement of the DPA and the IBA at 
$\sqrt{s_{\ga\ga}}\sim170\GeV$ 
both for semi-leptonic and for hadronic final states (in both cases 
the difference is well below 0.1\%) is of course accidental.
For the leptonic final state the difference is about 0.7\%.

As explained in \refse{se:IBA}, the intrinsic uncertainty of the IBA
is about $\sim\pm 2\%$, while the DPA accuracy is up to $\lsim0.5\%$ 
where it is applicable. 
Since the convolution of the hard $\ga\ga$ cross section,
in general, involves both the IBA (in the low-energy tail) and the DPA
(for $\sqrt{s_{\ga\ga}}>170\GeV$), the uncertainty of our cross-section 
prediction is in the range $0.5{-}2\%$, depending on the contribution of the
IBA part to the full convolution. Denoting the IBA and DPA parts of the
full cross section as $\Delta\sigma_{\IBA}$ and $\Delta\sigma_{\DPA}$
(both including the corresponding lowest-order contribution, so that
$\Delta\sigma_{\IBA}+\Delta\sigma_{\DPA}=\si$),
we can estimate the theoretical uncertainty (TU) of the corrected
cross section $\si$ to
\beq
\mathrm{TU} = \frac{\Delta\sigma_{\IBA}}{\si} \times 2\% \;+\;
\frac{\Delta\sigma_{\DPA}}{\si} \times 0.5\%.
\label{eq:TU}
\eeq
Table~\ref{tab:TU} illustrates this estimate for a few CM energies
$\sqrt{s_{\Pe\Pe}}$ for $\ga\ga\to\Pne\Pep\Pd\Pubar$. 
\begin{table}
\centerline{
\begin{tabular}{|c|c|c|c|c|c|c|c|}
\hline
$\sqrt{s_{\Pe\Pe}}/\GeV$ & 200 & 240 & 260 & 280 & 300 & 500 & 1000
\\
\hline
TU & 2.0\% & 1.9\% & 1.3\% & 0.8\% & 0.7\% & 0.5\% & 0.5\%
\\
\hline
\end{tabular} }
\label{tab:TU}
\caption{Estimates of the TU \refeq{eq:TU} for the 
${\cal O}(\alpha)$-corrected cross section of $\ga\ga\to\Pne\Pep\Pd\Pubar$
at various CM energies $\sqrt{s_{\Pe\Pe}}$.}
\end{table}
For $\sqrt{s_{\Pe\Pe}}\lsim 230\GeV$ our prediction possesses a TU of
$\sim 2\%$, because it is mainly based on the IBA, but already
for $\sqrt{s_{\Pe\Pe}}\gsim 300\GeV$ ($500\GeV$) 
the IBA contribution is widely suppressed
so that the DPA uncertainty of $\lsim 0.7\%$ ($0.5\%$) 
sets the precision of our calculation. 
We note, however, that the overall uncertainty of our calculation
certainly becomes worse as soon as TeV energies for 
$\sqrt{s_{\ga\ga}}$ are dominating because of the relevance of
high-energy logarithms beyond ${\cal O}(\alpha)$.

In \reffi{fig:sqrts.nospec}(e) the comparison of 
the full correction with the IBA is shown for a Higgs mass of $\MH=170\GeV$.
The IBA includes the Higgs resonance via an effective coupling and 
reflects the shape of the resonance quite well. 

\begin{figure}
\setlength{\unitlength}{1cm}
\centerline{
\begin{picture}(7.7,8)
\put(-1.7,-14.5){\includegraphics{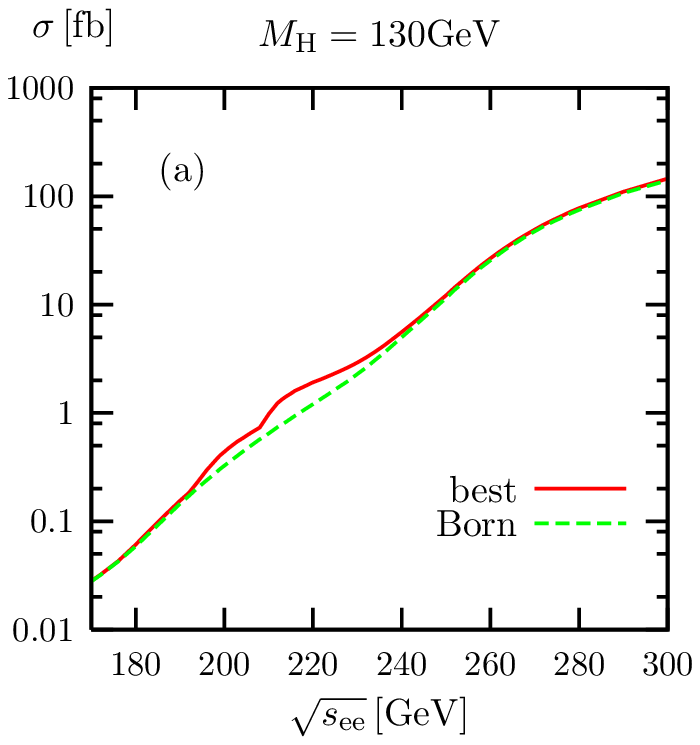}}
\end{picture}
\begin{picture}(7.5,8)
\put(-1.7,-14.5){\includegraphics{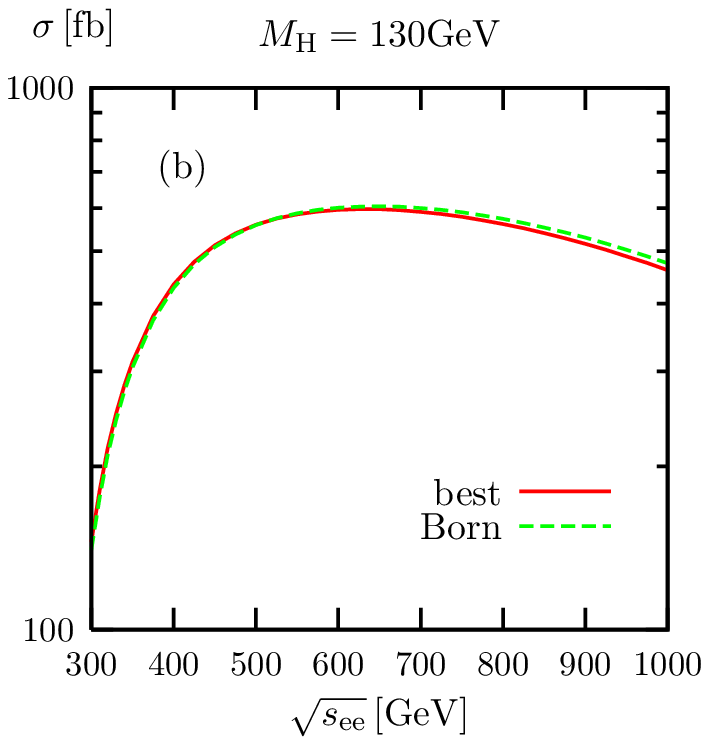}}
\end{picture} }
\centerline{
\begin{picture}(7.7,8)
\put(-1.7,-14.5){\includegraphics{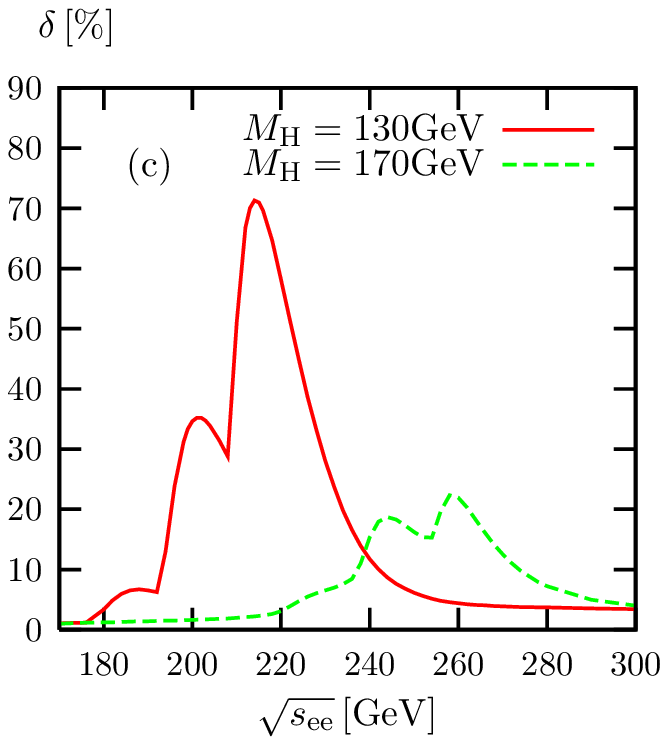}}
\end{picture}
\begin{picture}(7.5,8)
\put(-1.7,-14.5){\includegraphics{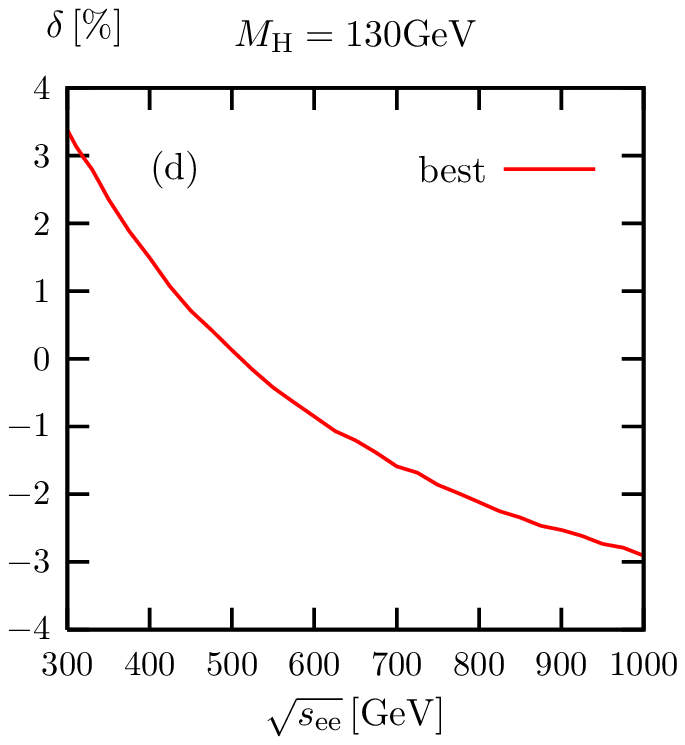}}
\end{picture} }
\caption{Integrated cross section for $\ga\ga\to\Pne\Pep\Pd\Pubar$
(upper plots) and relative radiative 
corrections (lower plots) including the convolution over the photon spectrum 
for Higgs masses of $\MH=130\GeV$ and $170\GeV$ (lower left plot).
For $\sqrt{s_{\Pe\Pe}}>300\GeV$ (shown on the r.h.s.) the ``best'' curve 
for $\MH=170\GeV$ practically coincides with the shown curve for
$\MH=130\GeV$.}
\label{fig:sqrts.spec}
\end{figure}

The cross section including the convolution over the photon spectrum 
as a function of CM energy is shown in \reffi{fig:sqrts.spec} 
for a Higgs mass of $\MH=130\GeV$ and in the lower left plot also 
for $\MH=170\GeV$. 
In the upper plots the integrated
cross sections are shown, and in the lower plots the corrections relative 
to the Born cross section. Recall that we use the IBA below 
$\sqrt{s_{\ga\ga}}=170\GeV$. This means, in particular, that the 
Higgs resonance is calculated from the effective coupling and 
not from the full DPA in this region. 
The interesting structure in the lower left plot 
reflects the shape of the photon spectrum convoluted with the Higgs resonance. 
Since the Higgs resonance 
is very narrow, a sizable contribution is only possible if
$x_1x_2s_{\Pe\Pe}\approx\MH^2$ where $x_1$ and $x_2$ are the energy fractions 
carried by the photons. The correction is 
very small at low $\sqrt{s_{\Pe\Pe}}$ where $x_1$ and $x_2$ 
have to be so large in order to match this condition that
the corresponding spectrum is extremely small. Increasing 
$\sqrt{s_{\Pe\Pe}}$ allows for lower values of $x_1$ and $x_2$.
For instance, for $\MH=130\GeV$,
the rise at $\sqrt{s_{\Pe\Pe}}\sim 180\GeV$ results from a region 
where both $x_1$ and $x_2$ are in the high-energy tail 
of the spectrum which is produced by multiple photon scattering. 
The peak at  $\sqrt{s_{\Pe\Pe}}\sim 200\GeV$ is caused by events where 
one photon comes from the high-energy tail and one from the dominant 
peak in the photon spectrum. Finally, at $\sqrt{s_{\Pe\Pe}}\gsim 210\GeV$ 
both $x_1$ and $x_2$ originate from 
the dominant photon-spectrum peak which causes the steep rise 
until $\sqrt{s_{\Pe\Pe}}\sim 220\GeV$.

\subsection{Distributions}

\begin{figure}
\setlength{\unitlength}{1cm}
\centerline{
\begin{picture}(7.7,8)
\put(-1.7,-14.5){\includegraphics{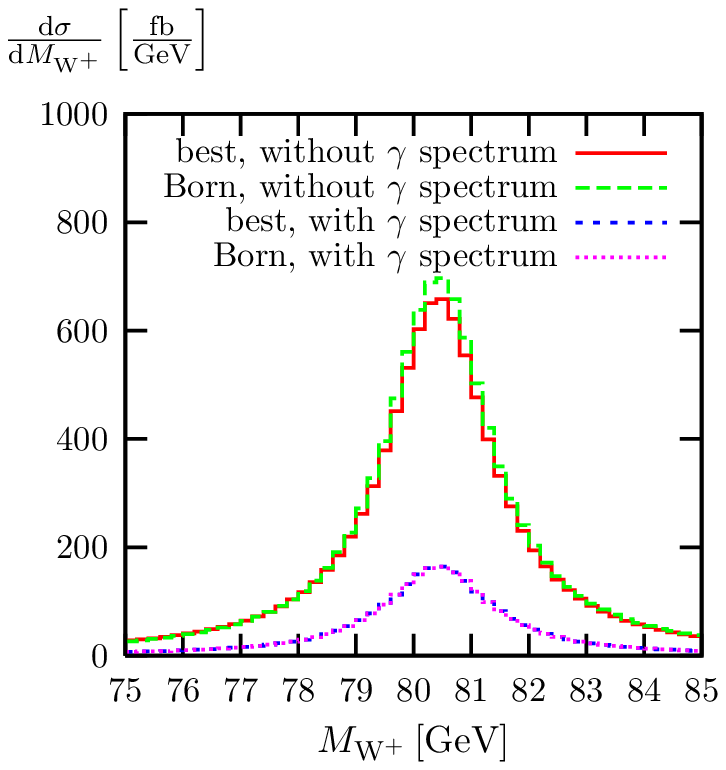}}
\end{picture}
\begin{picture}(7.5,8)
\put(-1.7,-14.5){\includegraphics{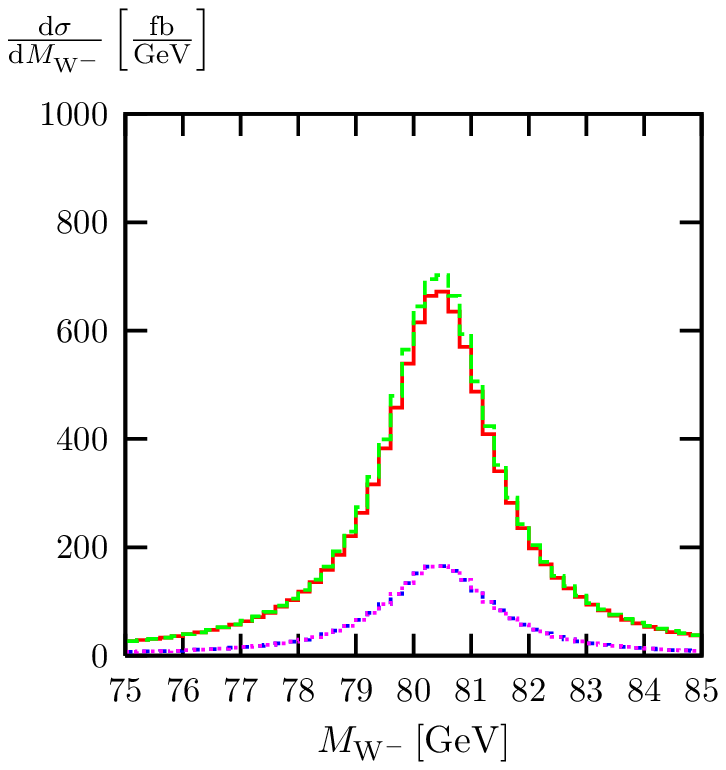}}
\end{picture}}
\centerline{
\begin{picture}(7.7,8)
\put(-1.7,-14.5){\includegraphics{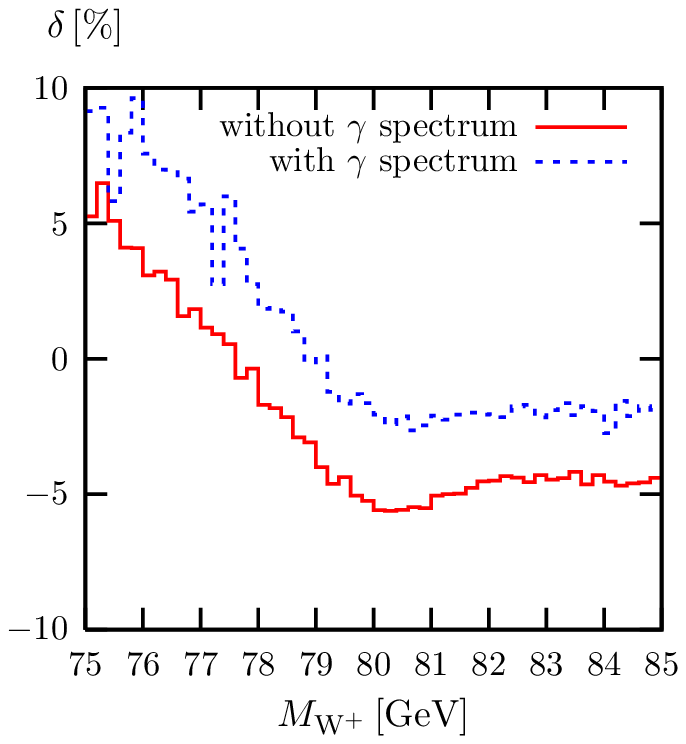}}
\end{picture}
\begin{picture}(7.5,8)
\put(-1.7,-14.5){\includegraphics{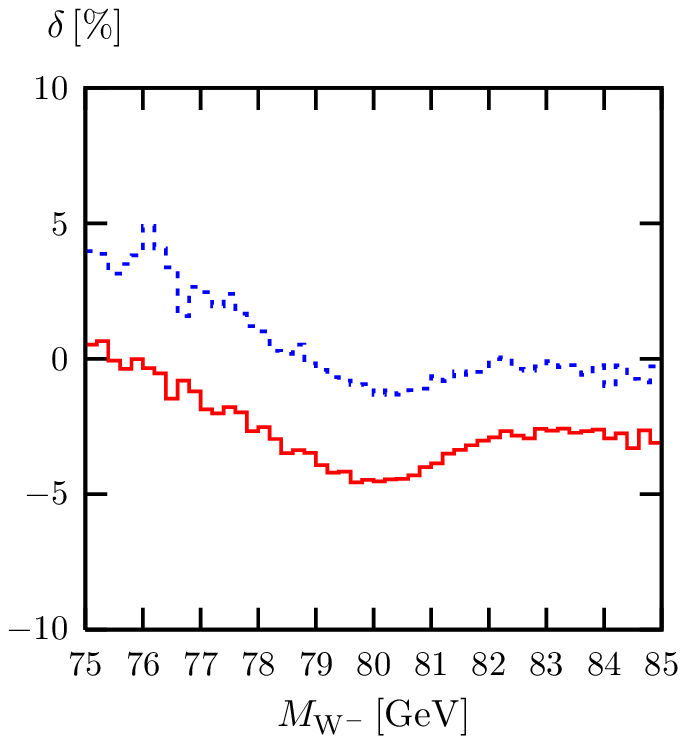}}
\end{picture}}
\caption{Invariant-mass distributions of the $\PW^+$ and $\PW^-$ bosons 
reconstructed from the $\Pne\Pep$ and $\Pd\Pubar$ pairs 
in the process $\ga\ga\to\Pne\Pep\Pd\Pubar$ at $\sqrt{s}=500 \GeV$.}
\label{fig:wmass}
\end{figure}

In \reffi{fig:wmass} we show the invariant-mass distributions for the 
$\Pne\Pep$ and $\Pd\Pubar$ pairs in the process $\ga\ga\to\Pne\Pep\Pd\Pubar$, 
both with and without convolution over the photon spectrum.
The upper plots show the absolute predictions, and the lower plots the 
corrections normalized to the Born predictions.   
Since we use $\sqrt{s_{\ga\ga}}=500\GeV$ or $\sqrt{s_{\Pe\Pe}}=500\GeV$, 
the corrections are shifted upwards when including the photon 
spectrum, because the effective energy of the photons is lower 
(cf.\ \reffi{fig:sqrts.nospec}). 
The shape of the corrections, however, 
is hardly changed by the convolution over the photon spectrum. 
As the shape of the corrections determine a possible shift of the peak of the 
invariant-mass distribution, it is of particular importance 
in the determination of the W-boson mass. The measurement of the 
W-boson mass can, e.g., be used for understanding and calibrating
the detector of a $\ga\ga$ collider. 

\begin{figure}
\setlength{\unitlength}{1cm}
\centerline{
\begin{picture}(7.7,8)
\put(-1.7,-14.5){\includegraphics{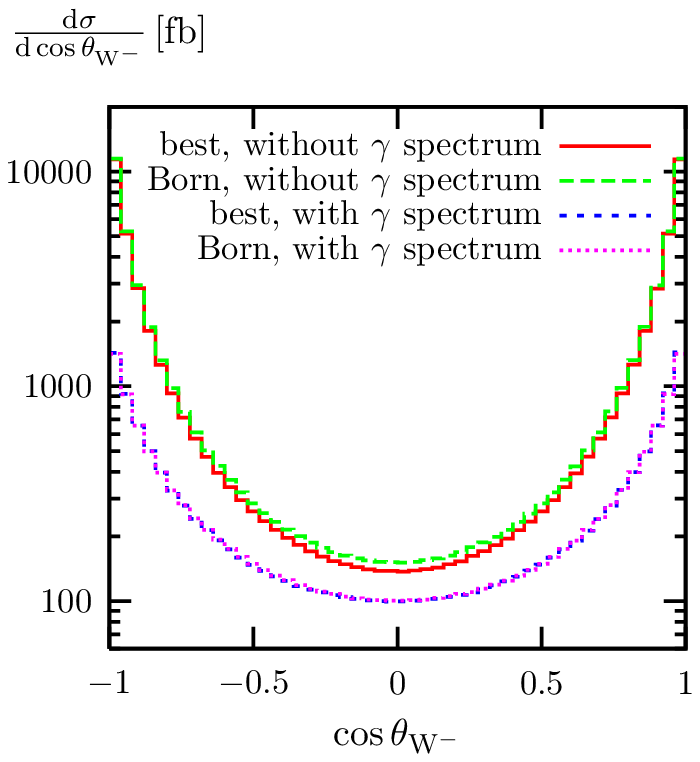}}
\end{picture}
\begin{picture}(7.7,8)
\put(-1.7,-14.5){\includegraphics{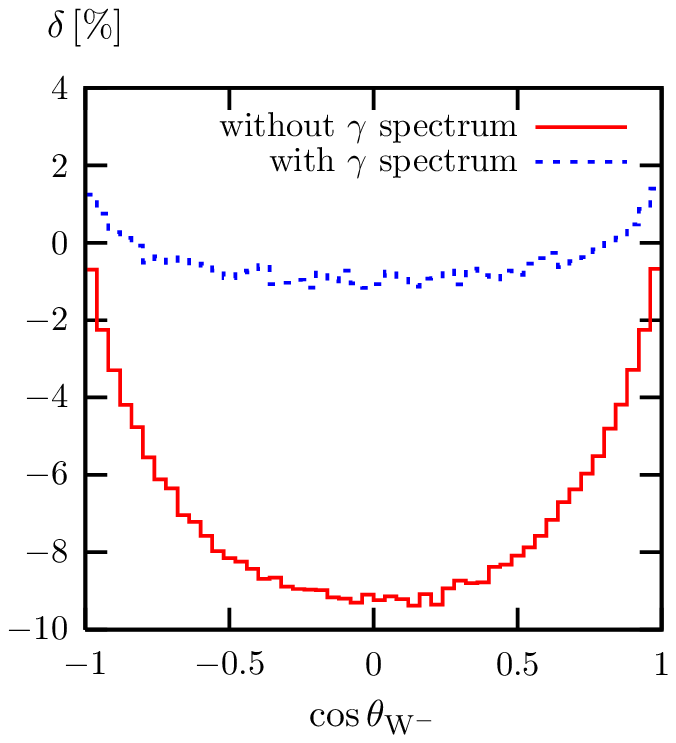}}
\end{picture}}
\caption{Angular distribution of the $\PW^-$ boson 
reconstructed from the $\Pd\Pubar$ pair
in the process $\ga\ga\to\Pne\Pep\Pd\Pubar$ at $\sqrt{s}=500\GeV$.}
\label{fig:wangle}
\end{figure}

The distribution in the W-boson production angle
is sensitive to anomalous couplings. 
In order to set bounds on these couplings it is mandatory to know 
radiative corrections,
because both anomalous couplings and radiative corrections typically
distort angular distributions.
The corresponding angular distribution of the $\Pd\Pubar$ 
system, which is equal within the statistical error 
to the distribution of the $\Pne\Pep$ system, 
is shown in \reffi{fig:wangle}. 
%The two possibilities with and without 
%the convolution over the photon spectrum are compared. 
While the correction without the photon spectrum is about $-9\%$
for W~bosons emitted perpendicular to the beam, the corrections are 
rather small when including the photon spectrum. As already explained above, 
the cross section is dominated by a region where the $\ga\ga$
CM energy is smaller. 
In fact, the relative correction $\de$ 
is accidentally small at $\sqrt{s_{\Pe\Pe}} \sim 500\GeV$ 
[cf.\ \reffi{fig:sqrts.spec}(d)]
and might also become larger if other cuts or event selection procedures
are applied.

\begin{figure}
\setlength{\unitlength}{1cm}
\centerline{
\begin{picture}(7.7,8)
\put(-1.7,-14.5){\includegraphics{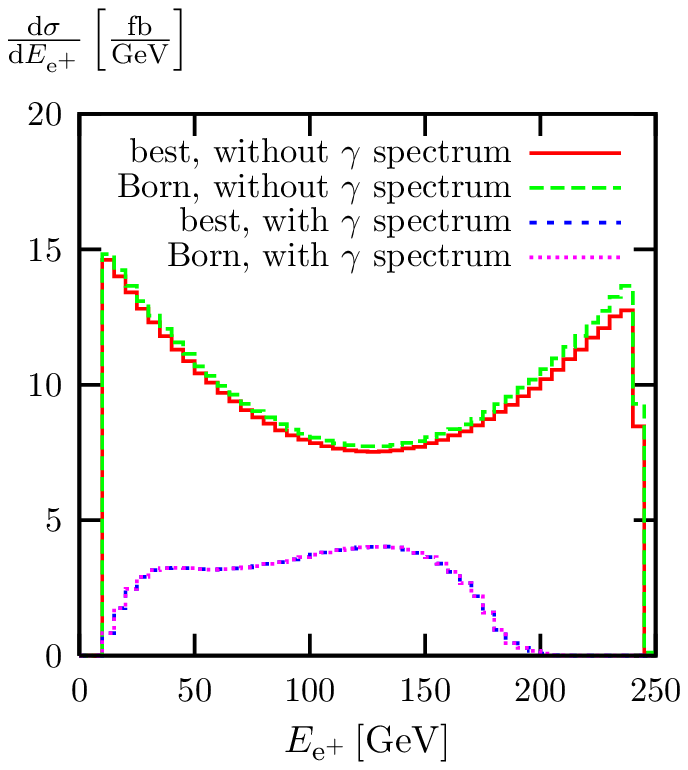}}
\end{picture}
\begin{picture}(7.5,8)
\put(-1.7,-14.5){\includegraphics{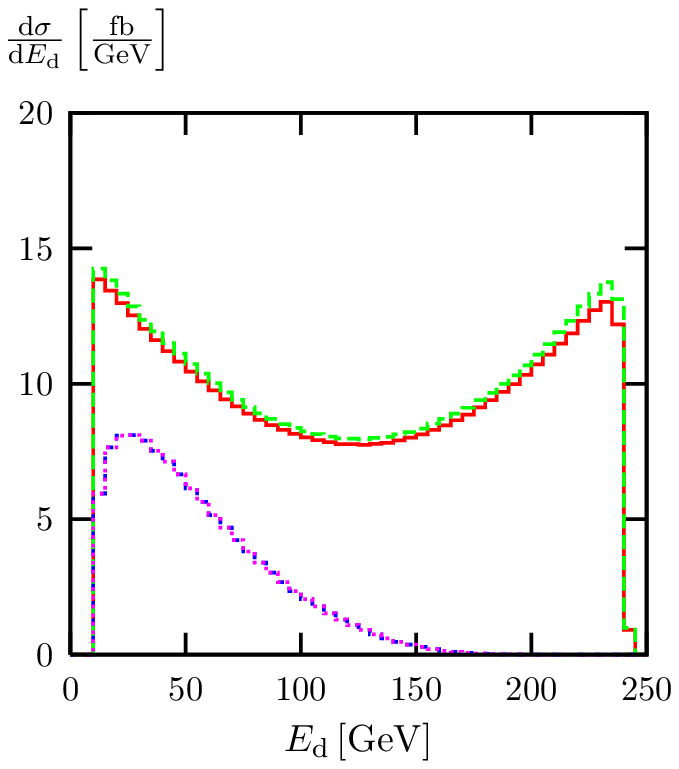}}
\end{picture}}
\centerline{
\begin{picture}(7.7,8)
\put(-1.7,-14.5){\includegraphics{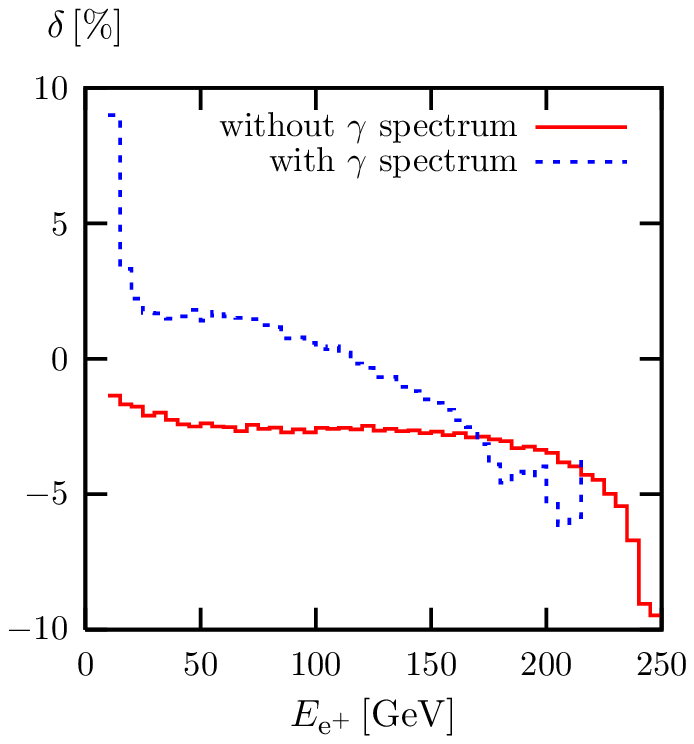}}
\end{picture}
\begin{picture}(7.5,8)
\put(-1.7,-14.5){\includegraphics{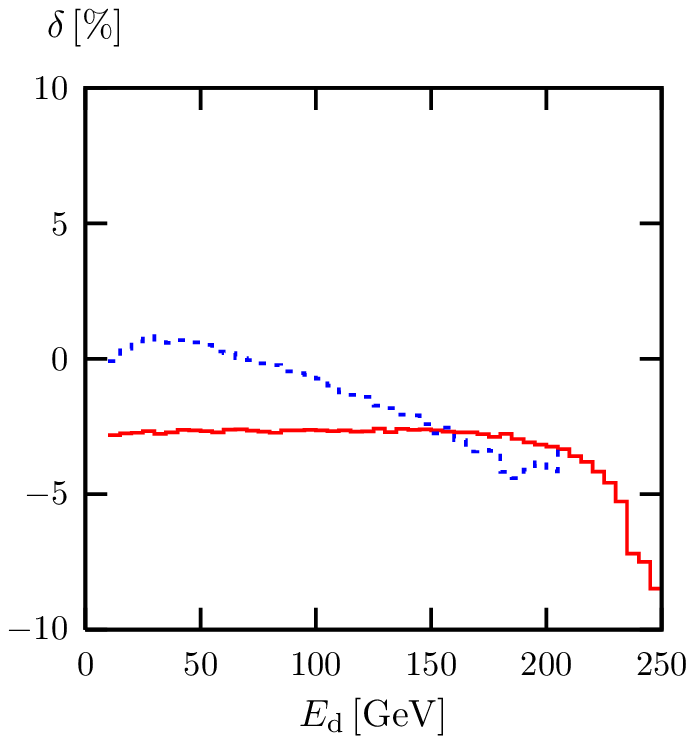}}
\end{picture}}
\caption{Energy distribution of $\Pep$ and $\Pd$ in the process 
$\ga\ga\to\Pne\Pep\Pd\Pubar$ at $\sqrt{s}=500\GeV$.}
\label{fig:energy}
\end{figure}

\begin{figure}
\setlength{\unitlength}{1cm}
\centerline{
\begin{picture}(7.7,8)
\put(-1.7,-14.5){\includegraphics{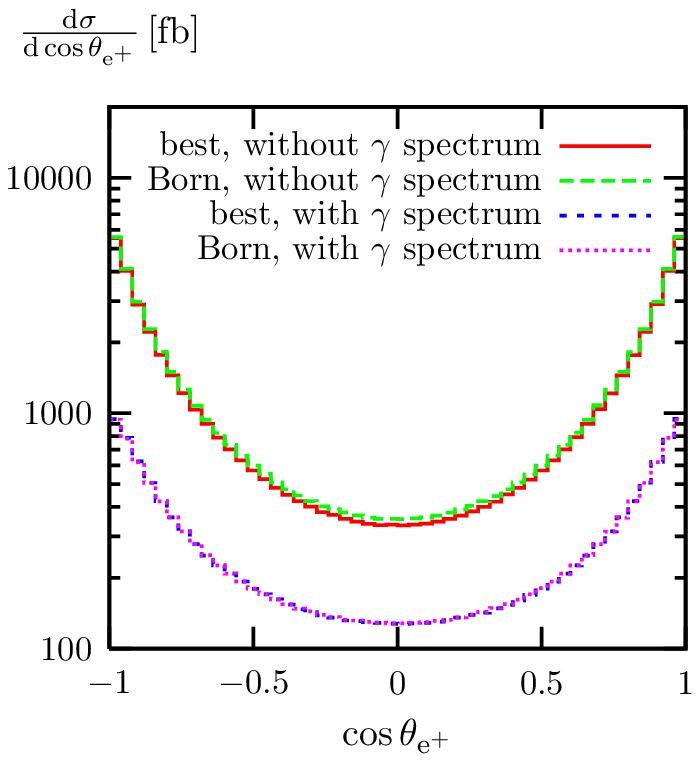}}
\end{picture}
\begin{picture}(7.5,8)
\put(-1.7,-14.5){\includegraphics{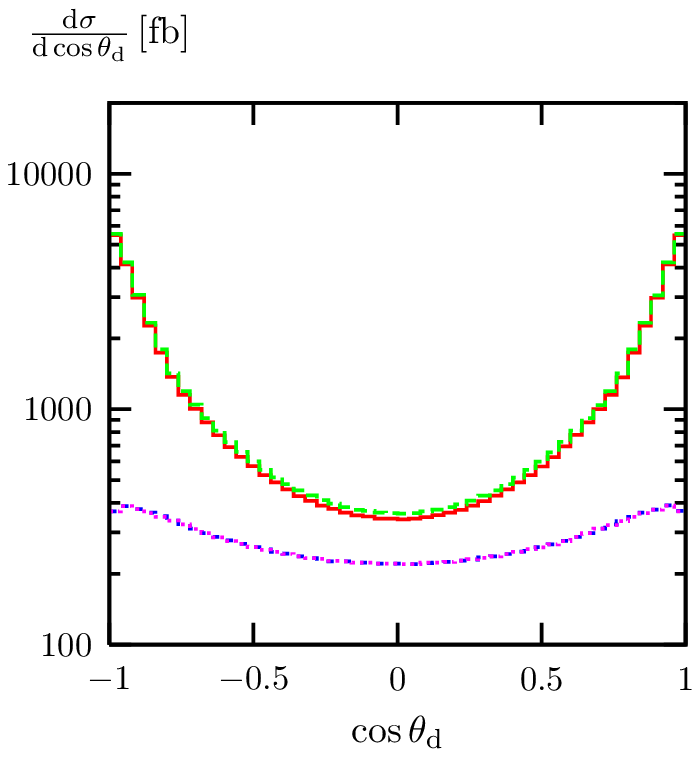}}
\end{picture}}
\centerline{
\begin{picture}(7.7,8)
\put(-1.7,-14.5){\includegraphics{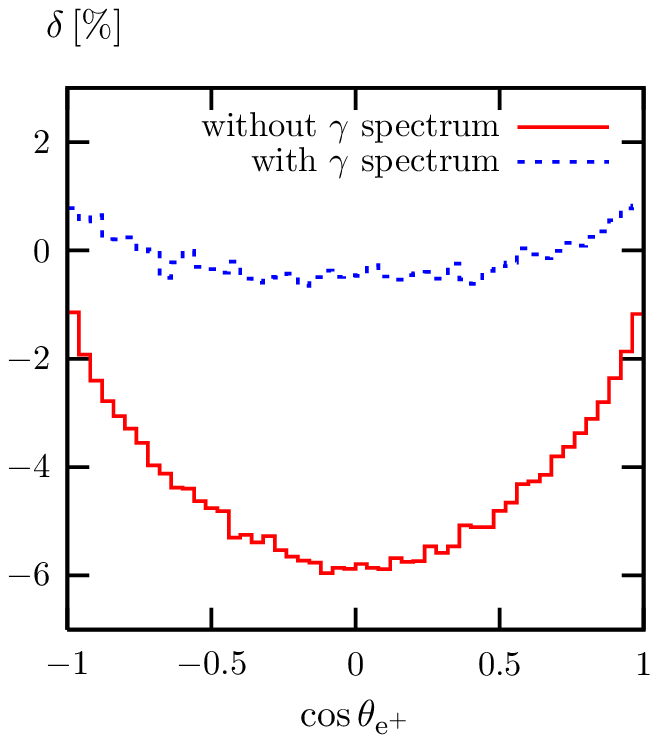}}
\end{picture}
\begin{picture}(7.5,8)
\put(-1.7,-14.5){\includegraphics{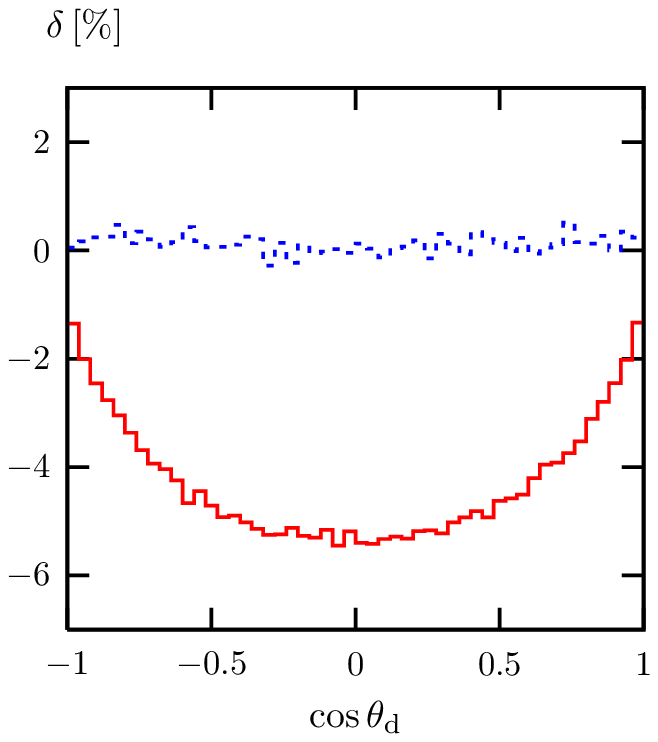}}
\end{picture}}
\caption{Distributions in the production angles of $\Pep$ and $\Pd$ 
in the process $\ga\ga\to\Pne\Pep\Pd\Pubar$ at $\sqrt{s}=500\GeV$.}
\label{fig:angle}
\end{figure}

Figure~\ref{fig:energy} 
shows the energy distribution of $\Pep$ and $\Pd$ 
for the process $\ga\ga\to\Pne\Pep\Pd\Pubar$. 
The characteristics of 
the Born cross section, especially the influence of the effective 
polarization of the photons after Compton backscattering, were 
explained in detail in Section 6.3 of \citere{Bredenstein:2004ef}. 
The relative corrections shown in the lower plots amount to a few per cent. 
For very low and very high energies, where the Born cross section is very 
small, the relative corrections 
in DPA are not reliable anymore. In this 
region the assumption that doubly-resonant diagrams dominate is not fulfilled.
The angular distributions for $\Pep$ and $\Pd$ 
are shown in \reffi{fig:angle}. 
The shape of the Born cross section and 
the influence of the photon spectrum were also explained in Section 6.3 of 
\citere{Bredenstein:2004ef}. Similar to the angular distributions of the 
$\Pne\Pep$ and $\Pd\Pubar$ systems, the corrections are maximal in 
a region where the fermions are emitted perpendicular to the beam. 
However, after including the photon spectrum, the corrections almost 
cancel as can be anticipated from \reffi{fig:sqrts.spec}(d) which 
shows that the corrections to the integrated cross section are 
almost zero at $\sqrt{s_{\Pe\Pe}}\sim500\GeV$.

\begin{figure}
\setlength{\unitlength}{1cm}
\centerline{
\begin{picture}(7.7,8)
\put(-1.7,-14.5){\includegraphics{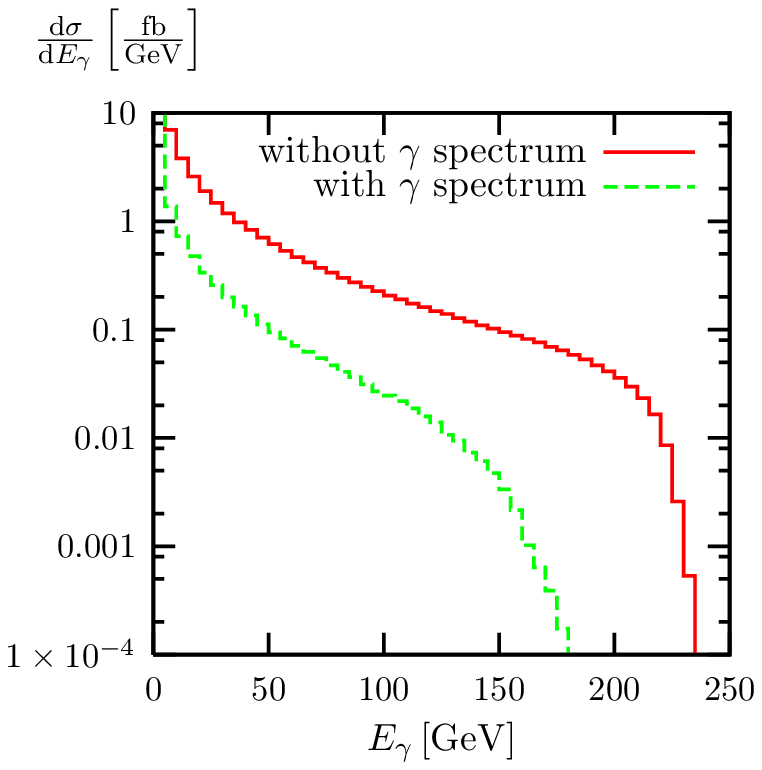}}
\end{picture}
\begin{picture}(7.5,8)
\put(-1.7,-14.5){\includegraphics{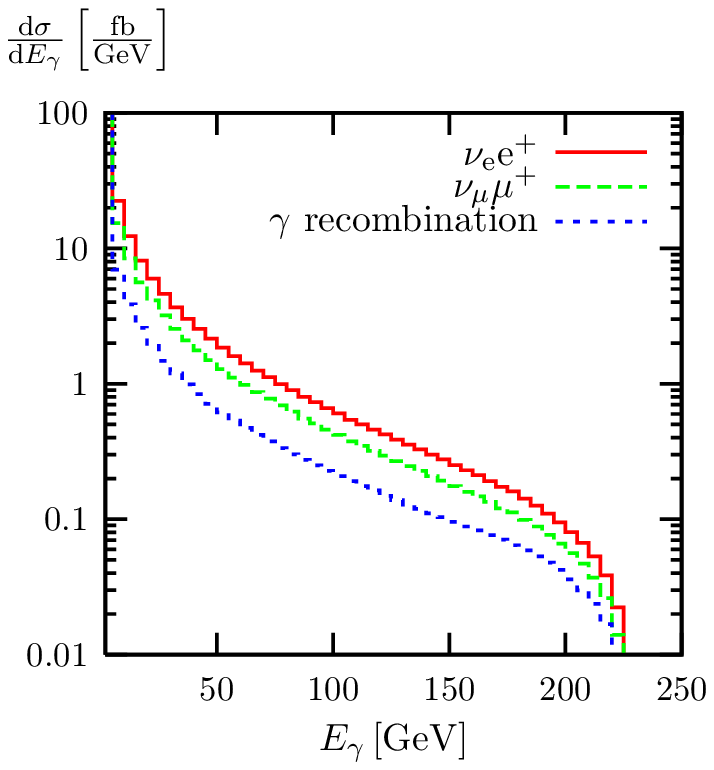}}
\end{picture}}
\caption{Energy distribution of the final-state photon   
in the processes $\ga\ga\to\Pne\Pep\Pd\Pubar+\gamma$
and $\ga\ga\to\nu_\mu\mu^+\Pd\Pubar+\gamma$ at $\sqrt{s}=500\GeV$.
The l.h.s.\ compares the distributions with and without convolution over the
photon spectrum (with photon recombination); the r.h.s.\
compares the cases with and without photon recombination
(without convolution over the photon spectrum).}
\label{fig:gamma}
\end{figure}

Finally, the energy distribution of the photon in the process 
$\ga\ga\to\Pne\Pep\Pd\Pubar{+}\ga$ is shown in \reffi{fig:gamma}. 
The distribution is dominated by the soft-photon pole at $k^0\to 0$ 
and decreases rapidly at higher energies. Comparing 
the distributions with and without convolution over the photon 
spectrum, the convolution shifts the curve to lower energies, because 
the initial-state photons already have less energy.

\subsection{Non-collinear-safe observables}
\label{se:num_noncollsafe}

\begin{figure}
\setlength{\unitlength}{1cm}
\centerline{
\begin{picture}(7.7,8)
\put(-1.7,-14.5){\includegraphics{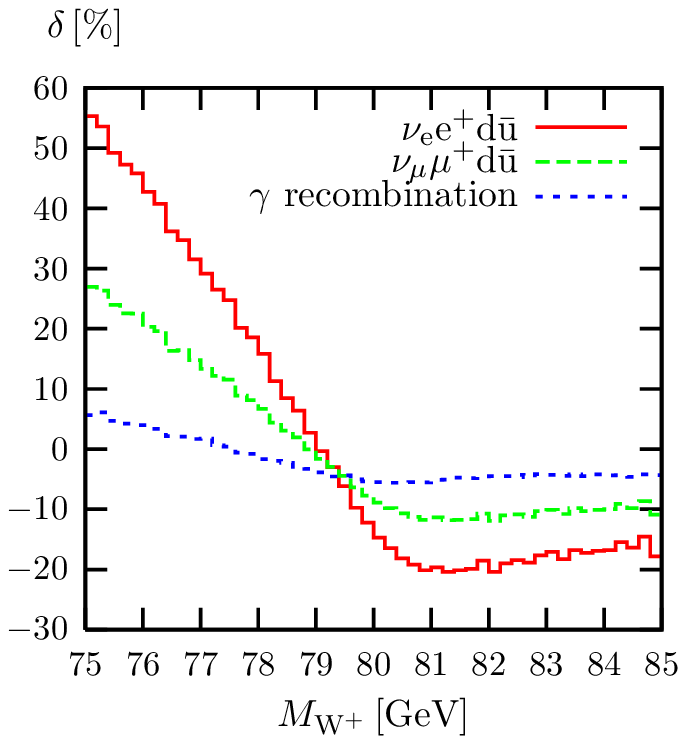}}
\end{picture}
\begin{picture}(7.5,8)
\put(-1.7,-14.5){\includegraphics{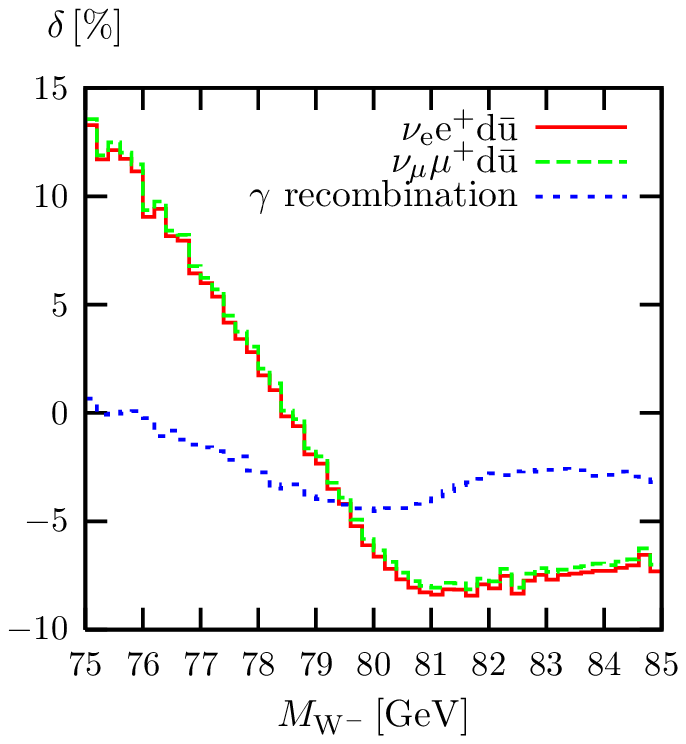}}
\end{picture} }
\caption{Invariant-mass distributions of the $\PW^+$ and $\PW^-$ boson 
reconstructed from the $\Pne\Pep$($\Pnmu\Pmup$) pair and 
$\Pd\Pubar$ pair in the process $\ga\ga\to\Pne\Pep\Pd\Pubar$
($\ga\ga\to\Pnmu\Pmup\Pd\Pubar$) at $\sqrt{s_{\ga\ga}}=500 \GeV$, with and 
without photon recombination.}
\label{fig:newsub.wmass}
\end{figure}

As explained in \refse{se:non-coll-safe},
the treatment of collinear singularities in 
non-collinear-safe observables deserves some care. 
Applying the generalizations of the subtraction and the slicing methods 
described above, we now turn to observables without 
photon recombination. Apart from that, the same phase-space cuts 
are applied as before. 
In \reffi{fig:newsub.wmass} we show the distributions of 
the $\Pne\Pep$, $\Pnmu\Pmup$, and $\Pd\Pubar$ pairs in the processes
$\ga\ga\to\Pne\Pep\Pd\Pubar$, $\Pnmu\Pmup\Pd\Pubar$. 
With photon recombination the leptonic invariant masses of the two
processes receive the same radiative 
corrections since the recombination guarantees the necessary inclusiveness 
so that all mass singularities cancel.
If the recombination is not applied, the distributions change drastically. 
Note, however, that the recombination is mainly a 
rearrangement of events, and omitting the recombination
affects the integrated cross section by less than $0.5\%$. 
With decreasing invariant masses the relative corrections rise,
while they are smaller at large invariant masses. 
The reason is that without recombination final-state radiation
(which is enhanced by mass logarithms) reduces the invariant mass of
the reconstructed W~boson, thereby shifting
events from the dominating resonant region to lower invariant mass values.
The recombination brings most of these events back to the resonance region,
because it prevents momentum loss from final-state radiation.
The l.h.s.\ of \reffi{fig:newsub.wmass} also shows a hierarchy in the
mass effects of the outgoing leptons as the slope for the 
$\Pne\Pep$ pair is much steeper than the slope for the $\Pnmu\Pmup$ pair 
due to the smaller mass of $\Pep$. The plot on the r.h.s.\ 
shows that the corrections for the $\Pd\Pubar$ pair are not as large as 
for the $\Pnmu\Pmup$ pair on the l.h.s., because the remaining 
mass terms behave like $Q_f^2\ln{m_f}$, 
where $Q_f$ denotes the charge of the fermion $f$. 
We also note that the corrections are smallest in the case with 
photon recombination because of the cancellation of 
all mass singularities.

\begin{figure}
\setlength{\unitlength}{1cm}
\centerline{
\begin{picture}(7.7,8)
\put(-1.7,-14.5){\includegraphics{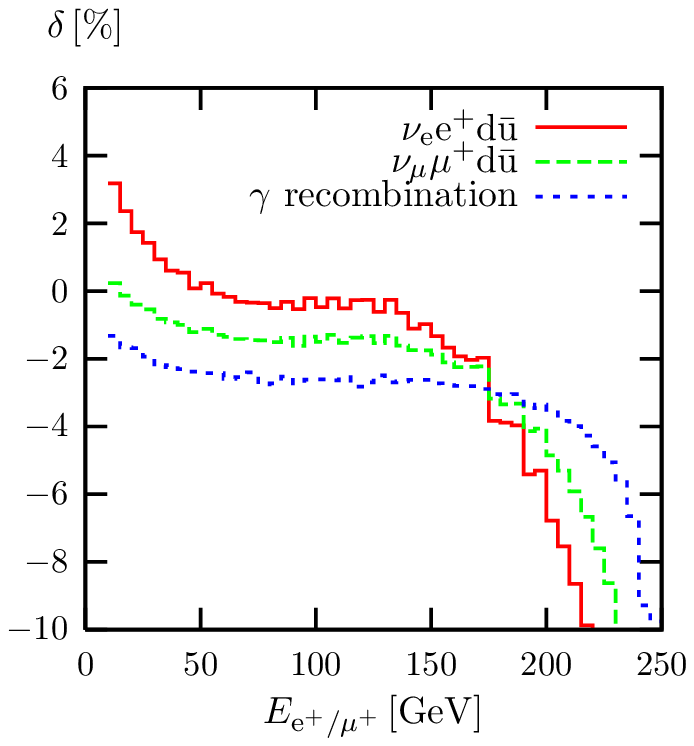}}
\end{picture}}
\caption{Energy distributions of $\Pep$ and $\Pmup$ in the processes 
$\ga\ga\to\Pne\Pep\Pd\Pubar$ and $\ga\ga\to\Pnmu\Pmup\Pd\Pubar$ 
at $\sqrt{s_{\ga\ga}}=500\GeV$, with and without photon recombination.}
\label{fig:newsub.energy}
\end{figure}

The photon recombination also affects the energy distributions 
of the fermions. Figure~\ref{fig:newsub.energy} shows this distribution 
for $\Pep$ and $\Pmup$ in the processes $\ga\ga\to\Pne\Pep\Pd\Pubar$ 
and $\ga\ga\to\Pnmu\Pmup\Pd\Pubar$ 
with and without recombination. 
In the former case the curves coincide, as explained above. 
The recombination attributes the photon to a fermion and, thus, 
shifts events to higher energies of the fermion. 
The mass-singular effect, which appears without recombination,
is again larger for $\Pep$ than for $\Pmup$.

The effect of the photon recombination on the photon-energy spectrum 
is shown in \reffi{fig:gamma}. Without recombination the distribution 
is shifted to higher photon energies because the recombination transfers 
events to the bin with zero photon energy. 
The difference is again bigger for the process 
$\ga\ga\to\Pne\Pep\Pd\Pubar$ than for $\ga\ga\to\Pnmu\Pmup\Pd\Pubar$, 
since the mass-singular logarithms of $\Pep$ are larger.

\section{Summary}
\label{se:summary}

In this paper we have described a calculation of
the ${\cal O}(\al)$ electroweak radiative corrections to $\ggwwffff$ in 
the electroweak Standard Model in the double-pole approximation (DPA). 
Technically, we follow the strategy of the {\sc RacoonWW} Monte Carlo
event generator for the corresponding $\Pep\Pem$ reaction.
This means, virtual corrections are treated in DPA and are decomposed
into factorizable and non-factorizable contributions,
while real-photonic corrections are based on complete lowest-order
matrix elements for $\ggffffg$. The combination of 
virtual and real corrections is done in two different ways:
by using the dipole subtraction method or by applying phase-space slicing.  

A detailed survey of numerical results for the ${\cal O}(\al)$ corrections
has been given, comprising results on integrated cross sections as well as 
angular, energy, and W-invariant-mass distributions.
In the W-pair threshold region the corrections are dominated by the
Coulomb singularity and are, thus, positive and of the order of a few
per cent. For increasing $\gamma\gamma$ scattering
energies the corrections become more and more
negative and reach about $-10\%$ in the TeV range for integrated
cross sections. For large scattering angles, where the Born cross section 
is relatively small, the impact of the corrections is usually larger.
Since the convolution with realistic photon beam spectra effectively
reduces the hard scattering energy, the size of the corrections is
usually somewhat reduced compared to the situation with monochromatic
photon beams. 
Typically, collinear-safe observables (i.e.\ where mass-singular
logarithms cancel due to sufficient inclusiveness) receive corrections
of a few per cent for energies of the $\Pem\Pem$ system before
Compton backscattering up to $1\TeV$.
As expected, non-collinear-safe observables receive very large corrections
(tens of per cent) because of the existence of logarithmic mass
singularities.

\begin{sloppypar}
The radiative corrections are implemented in a Monte Carlo generator
called {\sc Coffer$\gamma\gamma$}, which optionally includes anomalous
triple and quartic gauge-boson couplings in addition and performs a
convolution over realistic spectra of the photon beams.
The construction of this generator and lowest-order results obtained with
it have already been described in a previous publication 
\cite{Bredenstein:2004ef}.
At present, {\sc Coffer$\gamma\gamma$} is the only event generator that
includes both the decays of the W~bosons and radiative corrections,
thereby defining the state-of-the-art in the description of the 
processes $\ggwwffff(+\gamma)$.
\end{sloppypar}

\appendix
\section*{Appendix}
\renewcommand{\theequation}{A.\arabic{equation}}

\section{\boldmath{Transformation of the coefficient functions $F_j$}}
\label{app:sme}

In this appendix we describe the transformation of the coefficient 
functions $F_j$ for the factorizable virtual corrections \refeq{eq:fact2}
that transforms all $F_j$ into the helicity amplitudes of the on-shell 
process $\ggtoww$. 

The 36 SME $\M^{\ggww}_j$ of \citere{Denner:1995jv}, which fix the
coefficient functions $F_j$ by Eq.~\refeq{eq:decomp}, 
are defined for 36 different helicity configurations
which can be enumerated with a single index $l$,
\beq
\M^{\ggww}_j(\la_1,\la_2;\la_+,\la_-)\equiv M_{jl},
\quad l=(\la_1,\la_2,\la_+,\la_-),
\eeq
where $j,l=1,\dots,36$. The 36$\times$36 matrix $M$ is explicitly obtained
by inserting momenta and polarization vectors into 
the 36 independent SME $\M^{\ggww}_j$ of the 83 structures defined in
Eqs.~(5)--(9) of \citere{Denner:1995jv}.

If we transform the $F_j$ according to
\beq
\hat F_l=\sum_{j=1}^{36}F_j M_{jl},
\label{eq:newfj}
\eeq
the function $\hat F_l$ is the helicity amplitude for the on-shell process
$\ggtoww$ corresponding to the helicity configuration
$l=(\la_1,\la_2,\la_+,\la_-)$.
As such, it can be well approximated by the generalized Fourier series 
described in \refse{se:virtualeval}.
It is important to notice that in \citere{Denner:1995jv} the scattering
plane spanned by the beam axes and the produced W~bosons was rotated into the
$(x^1,x^3)$-plane, so that the SME $\M^{\ggww}_j$ depend only on
$s$ and $\cos\theta$, or equivalently on $s$ and $\hat t$.
Since, thus, the matrix $M$ is a function of $s$ and $\hat t$, also the
new functions $\hat F_l$ depend only on $s$ and $\hat t$, but not on the
azimuthal angle of the scattering plane or other on
kinematical variables.
According to Eq.~\refeq{eq:newfj}, the SME $\M_j$ transform as 
\beq
\hat\M_l=\sum_{j=1}^{36}(M^{-1})_{lj}\M_j,
\label{eq:newsme}
\eeq
where $M^{-1}$ denotes the inverse matrix of $M$.
By construction, the transformation decouples the different helicity
channels of $\ggtoww$. 
When including the W~decays in the SME, as done in Eq.~\refeq{eq:newsme},
this decoupling is somewhat disguised for the W-boson polarizations,
but still valid for the photon helicities. This means that the new
SME $\hat\M_l$ consist of four subsets, each of which contributes only
for one of the four different polarization combinations $(\la_1,\la_2)$
of the photons. 
In practice, we have evaluated and simplified the matrix $M$ and the 
new SME $\hat\M_l$ analytically as much as possible. 

\section{Dipole subtraction for non-collinear-safe photonic
final-state radiation}
\label{app:sub}

In \refse{se:dps-ncs} we have collected the relevant formulas for the
generalization of the dipole subtraction method to non-collinear-safe
observables. Here we describe the details of their derivation.
Specifically, we focus on the situation of light charged particles 
in the final state only; the more general case of massive particles 
and of charged particles in the initial state will be worked out elsewhere
\cite{gendpf}.
Although not made explicit in the main text, we keep track of the
polarizations of the outgoing particles.

\paragraph{Subtraction of singularities}

Generically the schematic form
of the subtraction procedure to integrate the squared matrix
element $\sum_{\lambda_\gamma} |\M_1|^2$ 
(summed over the photon polarizations $\la_\ga$) 
for real photon radiation
over the $(N+1)$-particle phase space $\rd\Phi_1$ reads
\beq
\int\rd\Phi_1\, \sum_{\lambda_\gamma} |\M_1|^2 =
\int\rd\Phi_1\, \Biggl(\sum_{\lambda_\gamma}|\M_1|^2-|\M_\sub|^2\Biggr)
+ \int\rd\tilde\Phi_0\,\otimes\left(\int [\rd k] \, |\M_\sub|^2\right),
\label{eq:sub}
\eeq
where $\rd\tilde\Phi_0$ is a phase-space element of the corresponding
non-radiative process and $[\rd k]$ includes the photonic phase space
that leads to the soft and collinear singularities.
The sign ``$\otimes$'' indicates that this factorization, in general, is 
not an ordinary product, but may contain also summations and convolutions.
The two contributions involving the subtraction function $|\M_\sub|^2$
have to cancel each other, however, they will be evaluated separately.
The subtraction function is constructed in such a way that
the difference $\left(\sum_{\lambda_\gamma}|\M_1|^2-|\M_\sub|^2\right)$
can be safely integrated over $\rd\Phi_1$ numerically and that the 
singular integration of $|\M_\sub|^2$ over $[\rd k]$ can be carried out
analytically, followed by a safe numerical integration over $\rd\tilde\Phi_0$.

In the dipole subtraction formalism, the subtraction function is
given by \cite{Dittmaier:1999mb,Roth:1999kk}
\beq
|\M_\sub(\Phi_1)|^2 =
-\sum_{i\neq j} Q_i \sigma_i Q_j \sigma_j e^2
\gsub_{ij,\tau}(p_i,p_j,k)
\left|\M_0\left(\tilde\Phi_{0,ij};\tau\kappa_i\right)\right|^2,
\label{eq:m2sub}
\eeq
where the sum runs over all emitter--spectator pairs $ij$, which are called
dipoles. Recall that both $i$ and $j$ are final-state particles in our case.
The relative charges are denoted $Q_f$ $(f=i,j)$, and
the sign factors $\si_f=\pm1$ correspond to the charge flow
($\sigma_f=+1$ for anti-fermions, $\sigma_f=-1$ for fermions).
The summation over $\tau=\pm1$ accounts for a possible flip in the
helicity $\kappa_i$ of the emitter $i$.
The singular behaviour of the subtraction function is contained in
the radiator functions $\gsub_{ij,\tau}(p_i,p_j,k)$, which depend
on the emitter, spectator, and photon momenta $p_i$, $p_j$, and $k$,
respectively. 
In the limit of small fermion masses the functions $\gsub_{ij,\tau}$
are related to the function $\gsub_{ij}$ of Eq.~\refeq{eq:gsubij}
for the unpolarized case by
$\gsub_{ij,+}=\gsub_{ij}$, $\gsub_{ij,-}=0$.
The squared lowest-order matrix element $|\M_0|^2$ 
of the corresponding non-radiative process enters the subtraction function
with modified emitter and spectator momenta $\tilde p_i$, $\tilde p_j$, 
as defined in Eq.~\refeq{eq:ptilde}.
The momenta are related by $p_i+p_j+k=\tilde p_i +\tilde p_j$, where
all the other particle momenta $p_k$ and $\tilde p_k$,
entering $|\M_1|^2$ and $|\M_0|^2$, respectively,
are the same, $\tilde p_k=p_k$. 
The modified momenta are constructed in such a way that
$\tilde p_i \to p_i+k$ in the collinear limit $(p_i k\to0)$.
Since we deal with light external fermions only,
we set all masses $m_f$ of external fermions to
zero whenever possible. This means that
$m_f=0$ can be consistently used in the integral 
$\int\rd\Phi_1\, \left(\sum_{\lambda_\gamma}|\M_1|^2-|\M_\sub|^2\right)$,
but that the readded contribution $\int [\rd k] \, |\M_\sub|^2$ in
general contains mass-singular terms of the form $\alpha\ln m_f$.

\begin{sloppypar}
In {\it collinear-safe}\ observables, 
and only those are considered for light fermions
in \citeres{Dittmaier:1999mb,Roth:1999kk}, a collinear fermion--photon system is
treated as one quasi-particle, i.e., in the limit where a charged fermion $i$ 
and $\gamma$ 
become collinear only the sum $p_i+k$ enters the procedures of implementing 
phase-space selection cuts or of sorting an event into a histogram bin
of a differential distribution. Technically this level of inclusiveness
is reached by {\it photon recombination}, a procedure that assigns the
photon to the nearest charged particle if it is close enough to it.
Of course, different variants for such an algorithm are possible,
similar to jet algorithms in QCD.
The recombination guarantees that for each photon radiation cone
around a charged particle $i$ the energy fraction of Eq.~\refeq{eq:zi},
$z_i=p_i^0/(p_i^0+k^0)$,
is fully integrated over. According to the KLN theorem \cite{Kinoshita:1962ur},
no mass singularity connected with final-state radiation remains.
Collinear safety facilitates the actual application of the subtraction
procedure as indicated in Eq.~\refeq{eq:sub}. In this case the
events resulting from the contributions of $|\M_\sub|^2$ can be 
consistently regarded as $N$-particle final states of the
non-radiative process with particle momenta as going into 
$\left|\M_0\left(\tilde\Phi_{0,ij}\right)\right|^2$, i.e.\
the emitter and spectator momenta are given by $\tilde p_i$,
$\tilde p_j$, respectively.
Owing to $\tilde p_i \to p_i+k$ in the collinear limits, the difference
$\left(\sum_{\lambda_\gamma}|\M_1|^2-|\M_\sub|^2\right)$ can be integrated
over all collinear regions, because all events that differ only in the value
of $z_i$ enter cuts or histograms in the same way.
The implicit {\it full} integration over all $z_i$ in the collinear cones,
on the other hand, implies that in the analytical integration of
$|\M_\sub|^2$ over $[\rd k]$ the $z_i$ integrations can be carried out
over the whole $z_i$ range.
\end{sloppypar}

In {\it non-collinear-safe}\ observables, 
not all photons within arbitrarily narrow collinear cones around outgoing
charged particles are treated inclusively. For a fixed cone axis the
integration over the corresponding variable $z_i$ is constrained by
a phase-space cut or by the boundary of a histogram bin. Consequently,
mass-singular contributions of the form $\alpha\ln m_i$ remain
in the integral.
Technically this means that the information on the variables $z_i$
has to be exploited in the subtraction procedure of Eq.~\refeq{eq:sub}.
The variable that takes over the role of $z_i$ in the individual
dipole contributions in $|\M_\sub|^2$ is $z_{ij}$, as defined
in Eq.~\refeq {eq:yijzij}, because 
in the collinear limit it behaves as $z_{ij}\to z_i$.
Thus, the integral 
$\int\rd\Phi_1\, \left(\sum_{\lambda_\gamma}|\M_1|^2-|\M_\sub|^2\right)$ 
can be performed over the whole phase space if the events associated with
$|\M_\sub|^2$ are treated as $(N+1)$-particle events with momenta
$p_i=z_{ij} \tilde p_i$,
$p_j=\tilde p_j$, and $k=(1-z_{ij})\tilde p_i$.
This modification, in turn, requires a generalization in the evaluation
of the second subtraction term on the r.h.s.\ of Eq.~\refeq{eq:sub},
because now the integral over $z_{ij}$ implicitly contained in
$[\rd k]$ depends on the cuts that define the observable. 

\paragraph{Integration of singular contributions}

For a final-state emitter $i$ and a final-state spectator $j$ with
masses $m_i$ and $m_j$ the integral of $\gsub_{ij,\tau}(p_i,p_j,k)$
over $[\rd k]$ is proportional to 
\beq
\Gsub_{ij,\tau}(P_{ij}^2) = \frac{\bar P_{ij}^4}{2\sqrt{\lambda_{ij}}}
\int_{y_1}^{y_2} \rd y_{ij}\, (1-y_{ij})
\int_{z_1(y_{ij})}^{z_2(y_{ij})} \rd z_{ij} \,
\gsub_{ij,\tau}(p_i,p_j,k),
\eeq
where $P_{ij}^2=(\tilde p_i+\tilde p_j)^2$ and
the definitions of Section~4.1 of \citere{Dittmaier:1999mb} are used.
There the results for $\Gsub_{ij,\tau}(P_{ij}^2)$ with generic or light masses
are given in Eqs.~(4.10) and (3.7), respectively.
In order to leave the integration over $z_{ij}$ open, the order of the two
integrations has to be interchanged, and the integral solely taken over
$y_{ij}$ is needed,
\beq
\bcGsub_{ij,\tau}(P_{ij}^2,z_{ij}) = 
\frac{\bar P_{ij}^4}{2\sqrt{\lambda_{ij}}}
\int_{y_1(z_{ij})}^{y_2(z_{ij})} \rd y_{ij}\, (1-y_{ij})\,
\gsub_{ij,\tau}(p_i,p_j,k).
\label{eq:bcGsubij}
\eeq
Note that the function $\bcGsub_{ij,\tau}(P_{ij}^2,z)$ is not needed
for finite photon mass $\la$, because the soft singularity appearing
at $z\to1$ can be split off by employing a $[\dots]_+$ prescription
in the variable $z$,
\beq
\bcGsub_{ij,\tau}(P_{ij}^2,z) =
\Gsub_{ij,\tau}(P_{ij}^2) \de(1-z) 
+ \left[\bcGsub_{ij,\tau}(P_{ij}^2,z)\right]_+.
\eeq
This procedure shifts the soft singularity into the quantity
$\Gsub_{ij,\tau}(P_{ij}^2)$, which is already known from
\citere{Dittmaier:1999mb}. Moreover, the generalization to
non-collinear-safe integrals simply reduces to the extra term
$\left[\bcGsub_{ij,\tau}(P_{ij}^2,z)\right]_+$, which cancels out
for collinear-safe integrals where the full $z$-integration is carried
out.

In the limit $m_i\to0$ and for $m_j=\la=0$ the boundary of the $y_{ij}$ 
integration is given by
\beq
y_1(z) = \frac{m_i^2(1-z)}{P_{ij}^2 z}, \quad 
y_2(z) = 1, 
\eeq
and the functions relevant for the integrand $\gsub_{ij,\tau}$ behave as
(see Section~4.1 of \citere{Dittmaier:1999mb})
\beq
p_i k = \frac{P_{ij}^2}{2} y_{ij}, \quad
R_{ij}(y) = 1-y, \quad r_{ij}(y) = 1.
\eeq
The evaluation of Eq.~\refeq{eq:bcGsubij} becomes very simple and yields
\beqar
\bcGsub_{ij,+}(P_{ij}^2,z) &=&
P_{ff}(z)\,\biggl[ \ln\biggl(\frac{P_{ij}^2 z}{m_i^2}\biggr)-1\biggl]
+(1+z)\ln(1-z), \nn\\ 
\bcGsub_{ij,-}(P_{ij}^2,z) &=& 1-z,
\label{eq:bcGsubij2}
\eeqar
where $P_{ff}(y)$ is the splitting function of Eq.~\refeq{eq:Pff}.
Equation~\refeq{eq:bcGsubij2} 
is correct up to terms suppressed by factors of $m_i$.
For completeness, we repeat the form of the full integral
$\Gsub_{ij,\tau}(P_{ij}^2)$ in the case of light masses,
\beq
\Gsub_{ij,+}(P_{ij}^2) = {\cal L}(P_{ij}^2,m_i^2) - \frac{\pi^2}{3} + 1,
\qquad
\Gsub_{ij,-}(P_{ij}^2) = \frac{1}{2},
\label{eq:Gij0}
\eeq
with the auxiliary function ${\cal L}$ of Eq.~\refeq{eq:L}.
The results for the functions $\bcGsub_{ij}$ and
$\Gsub_{ij}$ for the unpolarized case, 
as given in Eqs.~\refeq {eq:bcGsub} and \refeq{eq:Gsub},
are obtained by summing over the variable
$\tau=\pm 1$ which accounts for the spin-flip,
\beq
\bcGsub_{ij} = \bcGsub_{ij,+} + \bcGsub_{ij,-}, \qquad
\Gsub_{ij} = \Gsub_{ij,+} + \Gsub_{ij,-}.
\eeq

Finally, we give the explicit form of the $ij$ contribution
$|\M_{\sub,ij}(\Phi_1)|^2$ to the phase-space integral of the 
subtraction function,
\beqar
\lefteqn{\int\rd\Phi_1\,|\M_{\sub,ij}(\Phi_1)|^2
= -\frac{\alpha}{2\pi} Q_i\sigma_i Q_j\sigma_j  \,
\int\rd\tilde\Phi_{0,ij}\,
|\M_0(\tilde p_i,\tilde p_j;\tau\kappa_i)|^2 }
\nn\\[.3em]
&& {} \times
\Biggl\{ \Gsub_{ij,\tau}(P_{ij}^2) \Theta\Big(\tilde\Phi_{0,ij}\Big)
%\nn\\[.3em]
%&& {} \qquad
+\int_0^1 \rd z\, 
 \left[\bcGsub_{ij,\tau}(P_{ij}^2,z)\right]_+ 
\Theta\Big(p_i=z\tilde p_i,k=(1-z)\tilde p_i,\{\tilde p_{k\ne i}\}\Big)
\Biggr\}.
\nn\\
\label{eq:zintij}
\eeqar
The arguments of the step functions $\Theta(\dots)$
indicate on which momenta phase-space cuts are imposed.

\end{document}